\documentclass{article}
\usepackage{amssymb}
\usepackage{amsmath,bm}

\title{Gravitational Self Force in a Schwarzschild Background and the Effective One Body Formalism}
\author{Thibault Damour}
\date{\it Institut des Hautes Etudes Scientifiques, 35, route de Chartres, 91440 Bures-sur-Yvette, France}

\begin{document}

\maketitle

\begin{abstract}
We discuss various ways in which the computation of conservative Gravitational Self Force (GSF) effects on a point mass moving in a Schwarz\-schild background can inform us about the basic building blocks of the Effective One-Body (EOB) Hamiltonian. We display the information which can be extracted from the recently published GSF calculation of the first-GSF-order shift of the orbital frequency of the last stable circular orbit, and we combine this information with the one recently obtained by comparing the EOB formalism to high-accuracy numerical relativity (NR) data on coalescing binary black holes. The information coming from GSF data helps to break the degeneracy (among some EOB parameters) which was left after using comparable-mass NR data to constrain the EOB formalism. We suggest various ways of obtaining more information from GSF computations: either by studying eccentric orbits, or by focussing on a special zero-binding zoom-whirl orbit. We show that logarithmic terms start entering the post-Newtonian expansions of various (EOB and GSF) functions at the fourth post-Newtonian (4PN) level, and we analytically compute the first logarithm entering a certain, gauge-invariant ``redshift'' GSF function (defined along the sequence of circular orbits).
\end{abstract}

\section{Introduction}

The detection of gravitational waves from coalescing binary systems depends upon the prior knowledge of accurate theoretical models of the emitted gravitational waveforms, so as to be able to extract the gravitational wave signal from the noisy output of the detector. There has been much progress, over the past few years, on the development of accurate computational tools for describing the motion and radiation of (comparable-mass) compact binary systems (i.e. systems made of black holes or neutron stars). These computational tools are based either on analytical methods, on numerical ones, or on various combinations of both. The first formalism which made several quantitative and qualitative predictions about the entire coalescence process of comparable-mass circularized black hole binaries (from early inspiral to late ringing) is the analytical Effective One Body (EOB) formalism \cite{Buonanno:1998gg,Buonanno:2000ef,Damour:2001tu}. [Note that the EOB formalism uses, as essential inputs, the results of high-order post-Newtonian (PN) expanded results (see \cite{Blanchet:2006zz} for a review). However, it does not use PN results in their original ``Taylor-expanded form'', but rather in some suitably {\it resummed} form.] Soon afterwards, a combination of (short) full numerical simulations, with a ``close limit approximation'' \cite{Price:1994pm} to the ringing final black hole led to the first, numerical-based, (approximate) description of the coalescence of (comparable mass) circularized black hole binaries \cite{Baker:2001nu,Baker:2001sf}. Recently, several breakthroughs in numerical relativity (NR) \cite{Pretorius:2007nq} have allowed numerical methods to describe, with very high accuracy, the motion and radiation of coalescing black holes. These impressive NR achievements do not, however, render obsolete the development of {\it analytical} methods for describing the motion and radiation of coalescing black holes. Indeed, in spite of the high computer power used in NR simulations, the calculation of one sufficiently long waveform (corresponding to specific values of the many continuous parameters describing the considered binary system) takes on the order of several weeks. For detection purposes, one needs to compute tens of thousands of theoretical templates, so as to densely sample the full parameter space. This is a clear motivation for developing accurate analytical models of waveforms. One avenue for doing so is to use the natural flexibility of the EOB formalism (which was emphasized early on \cite{Damour:2001tu}) to ``tune'' some of the theoretical EOB parameters representing yet uncalculated, higher-order effects until the EOB waveform ``best fits'' a sparse sample of high-accuracy NR waveforms. Over the last years, this strategy has been vigorously pursued and has led to an impressive analytical/numerical agreement, with residual differences in phase and amplitude on the order of the current numerical errors \cite{Buonanno:2007pf,Damour:2007yf,Damour:2007vq,Damour:2008te,Boyle:2008ge,Damour:2009kr,Buonanno:2009qa}. The most recent, and most accurate, implementation of this strategy used {\it only two EOB flexibility parameters}, denoted $a_5$ and $a_6$, and found that there is a strong degeneracy between $a_5$ and $a_6$ in the sense that one can find an excellent NR-EOB agreement within a long and thin banana-like region in the $(a_5 , a_6)$ plane \cite{Damour:2009kr,Damour:2009ic}. This ``good fit region'' approximately extends between the points $(a_5 , a_6) = (0,-20)$ and $(a_5 , a_6) = (-36,+520)$ \cite{Damour:2009ic}.

\smallskip

The main purpose of the present work is to study to what extent the computation of gravitational self force (GSF) effects on a point mass moving in a Schwarzschild background can inform us about some of the yet uncalculated higher-order theoretical EOB parameters, such as the just mentionned $a_5$ and $a_6$ parameters. The GSF program aims at describing the motion and radiation of a small compact object (of mass $m_1$) moving around (and eventually inspiralling into) a much larger central black hole (of mass $m_2$). By contrast to the computational tools mentionned above (PN, EOB, NR), which can study binary systems with arbitrary (symmetric) mass ratio\footnote{Note, however, that NR simulations become increasingly difficult and time consuming as the symmetric mass ratio $\nu = m_1 m_2 / (m_1 + m_2)^2$ becomes small.} $\nu \equiv m_1 m_2 / (m_1 + m_2)^2$, the GSF program is a priori limited to the extreme mass ratio case, $m_1 \ll m_2$, i.e. $\nu \ll 1$. Roughly speaking, the GSF program is motivated by the planned space-based interferometric gravitational wave detector LISA \cite{LISA}, while the other computational programs are motivated by the higher-frequency ground-based interferometric detectors (such as LIGO \cite{LIGO}). The GSF program has been under development for a long time (essentially since the classic work of Refs.~\cite{Regge:1957td,DeWitt:1960fc,DeWitt:1964de,Zerilli:1971wd,Davis:1972ud}), but has started to reach fruition only very recently. [See \cite{Barack:2009ux,Detweiler:2009ah} for recent reviews of the GSF program.] We have particularly in mind here the recent breakthrough of Barack and Sago (BS), Ref.~\cite{Barack:2009ey}, which computed the first-order GSF correction to the frequency of the Last (circular) Stable Orbit (LSO) of a small point mass $m_1$ orbiting around a large (non spinning) black hole of mass $m_2$. [Another interesting recent result is the computation of the first-order GSF correction to a certain gauge invariant function of the sequence of circular orbits \cite{Detweiler:2008ft,Blanchet:2009sd}.] We wish here to relate the published result of \cite{Barack:2009ey} to the parameters entering the EOB formalism (and notably the parameters $a_5 , a_6 , \ldots$ mentionned above), and suggest further ways of using the GSF program for learning more about the theoretical building blocks of the EOB formalism. Hopefully, such a bringing together of the GSF and EOB programs might benefit both, and might thereby help one to develop better theoretical models of coalescing binaries. Note finally that we shall only consider here the {\it conservative} aspects of both the EOB and the GSF. Indeed, the result of \cite{Barack:2009ey} only concerns the time-symmetric, conservative piece of the GSF. Note that the influence, due to conservative interactions, of a non vanishing symmetric mass ratio $\nu$ (with $0 \leq \nu \leq 1/4$) on the frequency of the LSO was first studied within the EOB formalism in \cite{Buonanno:1998gg} at the 2PN level, and in \cite{Damour:2000we} at the 3PN level. As for the influence of time-odd, radiative interactions on the location of the LSO it was first studied in \cite{Buonanno:2000ef} (for arbitrary $\nu \leq 1/4$) and shown there to lead to a ``blurring'' of the LSO frequency of fractional order $\nu^{2/5}$. [See also the work of \cite{Ori:2000zn}, restricted to the small mass ratio case $\nu \ll 1$.]

\section{Notation, choice of coordinate system, and physical units}
\setcounter{equation}{0}

Let us start by warning the reader about a conflict of notation, between the standard EOB notation and a usual GSF one, which can bring confusion. We shall here adhere to the standard EOB notation in which the two masses of the binary system are denoted $m_1$ and $m_2$ (with, say, $m_1 \leq m_2$ to fix ideas) and where one then defines
\begin{equation}
\label{eq2.1}
M \equiv m_1 + m_2 \, ; \quad \mu \equiv \frac{m_1 m_2}{m_1 + m_2} \, ; \quad \nu \equiv \frac{\mu}{M} = \frac{m_1 m_2}{(m_1 + m_2)^2} \, .
\end{equation}
By contrast, many GSF works use the letter $M$ to denote the large mass (denoted $m_2$ in our notation), and the letter $\mu$ to denote the small mass ($m_1$ in our notation)\footnote{We find helpful to use the mnemonics $1 < 2$ to remember that, by definiton, $m_1 < m_2$ (and $m_1 \ll m_2$ in the extreme mass ratio case).}.

\smallskip

Another possible source of confusion concerns the choice of coordinate system to describe the binary system. Though we shall only work with (formally) ``gauge invariant'' quantities, this still leaves some potentially confusing ambiguity. Indeed, in PN and EOB (as well as NR) studies, it is tacitly assumed that one uses an ``asymptotically flat coordinate system'', i.e. a coordinate system which exhibits, in a standard manner\footnote{Possibly after a trivial transformation from Minkowskian-like to other coordinates, say of the $(t,r,\theta,\varphi)$ type.}, the asymptotic flatness ($g_{\mu\nu} \to \eta_{\mu\nu}$) of the metric generated by the considered binary system. By contrast, while some GSF works (e.g. that of \cite{Detweiler:2008ft} based on a ``Regge-Wheeler gauge'') do use ``asymptotically flat'' coordinates, the GSF works that use a ``Lorenz gauge'' for studying perturbations $h_{\mu\nu} (x^{\lambda})$ off a Schwarzschild background $g_{\mu\nu}^{(0)} (x^{\lambda})$ (as is the case of the work \cite{Barack:2009ey} that we shall consider) actually use a coordinate system which is {\it not} (explicitly) ``asymptotically flat''. Indeed, it was shown by Detweiler and Poisson \cite{Detweiler:2003ci}, and by Barack and Lousto \cite{Barack:2005nr} (building up on results of Zerilli \cite{Zerilli:1971wd}) that the {\it unique} low multipole ($\ell \leq 2$) contributions to the first-order Lorenz gauge metric perturbation generated by a point mass in {\it circular} motion are such that the monopole ($\ell = 0$) contribution (and only the monopole) yields a metric perturbation which does not decay as $r \to \infty$, but tends to
\begin{eqnarray}
\label{eq2.2}
\lim_{r \to \infty} h_{\mu\nu}^{\rm Lorenz} (x^{\lambda}) \, dx^{\mu} dx^{\nu} 
&= &- \, 2 \, \frac{Gm_1 \, \hat E_1}{r_0 (1-2 G m_2 / r_0)} \, dt^2  \\
&+ &dr^2 + r^2 d\theta^2 + r^2 \sin^2 \theta d \varphi^2 + {\mathcal O} (m_1^2) \, . \nonumber 
\end{eqnarray}
Here, $r_0$ is the Schwarzschild (areal) radial coordinate of the considered circular orbit, and
\begin{equation}
\label{eq2.3}
\hat E_1 \equiv \frac{E_1}{m_1} \equiv - \, u_t^1 \, ,
\end{equation}
the $(m_1)$ particle's conserved energy per unit rest mass. Note that we generally use units such that $c=1$ (and often also $G=1$), and a ``mostly plus'' signature $(-+++)$. Adding the Lorenz-gauge perturbation (\ref{eq2.2}) to the (``asymptotically flat'') background Schwarzschild metric leads to a perturbed metric for the binary system such that
\begin{eqnarray}
\label{eq2.4}
\lim_{r \to \infty} (g_{\mu\nu}^{(0){\rm Schw}} + h_{\mu\nu}^{\rm Lorenz}) \, dx^{\mu} dx^{\nu} 
&= &- \, (1+2\alpha) \, dt_{\rm Lorenz}^2 \\
&+ &dr^2 + r^2 d\theta^2 + r^2 \sin^2 \theta d \varphi^2 + {\mathcal O} (m_1^2) \, , \nonumber 
\end{eqnarray}
where
\begin{equation}
\label{eq2.5}
\alpha \equiv \frac{Gm_1 \, \hat E_1}{r_0 (1-2 Gm_2 / r_0)} \, .
\end{equation}
In the case that we consider here of circular orbits (and that will suffice for our purpose), the explicit expression, in terms of the orbital radius $r_0$, of the specific conserved energy $\hat E_1$ is
\begin{equation}
\label{eq2.6}
\hat E_1 = \frac{1-2Gm_2 / r_0}{\sqrt{1-3 Gm_2 / r_0}} \, ,
\end{equation}
so that (\ref{eq2.5}) can be reexpressed as
\begin{equation}
\label{eq2.7}
\alpha = \frac{Gm_1}{r_0 \, \sqrt{1-3 Gm_2 / r_0}} \, .
\end{equation}
The result (\ref{eq2.4}) is not new, and played, in particular, a crucial role in a recent work \cite{Sago:2008is} which compared the numerical results of the GSF obtained by two different methods, using different gauges. We are spelling it out again explicitly here to emphasize the following point.

\smallskip

Through first order in $m_1$, we need to {\it renormalize} the time coordinate associated to Lorenz-gauge GSF calculations to work in normal, ``asymptotically flat coordinates''. Namely, in the case considered here of a point mass $m_1$ moving on a circular orbit of radius $r_0$, we need to introduce
\begin{equation}
\label{eq2.8}
t_{\rm flat} = (1 + \alpha + {\mathcal O} (m_1^2)) \, t_{\rm Lorenz} \, ,
\end{equation}
together with $x_{\rm flat} \equiv r \sin \theta \cos \varphi$, $y_{\rm flat} = r \sin \theta \sin \varphi$, $z_{\rm flat} = r \cos \theta$, to ensure that the perturbed metric $ds^2 = (g_{\mu\nu}^{(0)} + h_{\mu \nu}) \, dx^{\mu} dx^{\nu}$ tends, as $r \to \infty$, towards the standard Poincar\'e-Minkowski metric $\underset{r \to \infty}{\lim} ds^2 = -dt_{\rm flat}^2 + dx_{\rm flat}^2 + dy_{\rm flat}^2 + dz_{\rm flat}^2$. It is a standard practice in relativistic gravity to use such ``asymptotically flat coordinates'' because they relate coordinates to {\it physical units}. For instance, the asymptotically measured orbital frequency of, say, a circular orbit, expressed in the physical time units (say the SI second) defined by the metric $ds^2$, $\Omega_{\rm phys}$, coincides with the coordinate angular frequency defined in ``flat'' coordinates $\Omega_{\rm flat} = d\varphi / dt_{\rm flat}$. For the purpose of gravitational wave observations, we are interested in the value of such physical-units frequencies: $\Omega_{\rm phys} = \Omega_{\rm flat} = d\varphi / dt_{\rm flat}$. In conclusion, the physical-unit value of a particular frequency $\Omega_{\rm phys} = \Omega_{\rm flat}$ differs from the corresponding Lorenz-gauge frequency $\Omega_{\rm Lorenz} = d\varphi / dt_{\rm Lorenz}$ by the ``renormalization'' factor deduced from (\ref{eq2.8})
\begin{equation}
\label{eq2.9}
\Omega_{\rm phys} = \frac{d\varphi}{dt_{\rm flat}} = \frac{1}{(1+\alpha +{\mathcal O} (m_1^2))} \, \frac{d\varphi}{dt_{\rm Lorenz}} = (1 - \alpha + {\mathcal O} (m_1^2)) \, \Omega_{\rm Lorenz} \, .
\end{equation}

Applying Eq.~(\ref{eq2.9}) to the recent, Lorenz-gauge GSF computation of the LSO frequency \cite{Barack:2009ey}, namely
\begin{equation}
\label{eq2.10}
Gm_2 \, \Omega_{\rm Lorenz}^{\rm LSO} = 6^{-3/2} \left[ 1 + c_{\Omega}^{\rm BS} \, \frac{m_1}{m_2} + {\mathcal O} \left( \left( \frac{m_1}{m_2} \right)^2 \right) \right] \, ,
\end{equation}
with 
\begin{equation}
\label{eq2.11}
c_{\Omega}^{\rm BS} = 0.4870(6) \, ,
\end{equation}
(where the number in parenthesis indicates the error on the last digits), we conclude that the physical-unit LSO frequency is
\begin{equation}
\label{eq2.12}
G m_2 \, \Omega_{\rm phys}^{\rm LSO} = 6^{-3/2} \left[ 1 + c_{\Omega}^{\rm BS} \, \frac{m_1}{m_2} - \alpha^{\rm LSO} + {\mathcal O} \left( \left( \frac{m_1}{m_2} \right)^2 \right) \right] \, ,
\end{equation}
where $\alpha^{\rm LSO}$ is the value of $\alpha$, Eq.~(\ref{eq2.7}), at the (zeroth order) LSO, i.e. $r_0 = 6 G m_2 + {\mathcal O} (m_1)$, so that
\begin{equation}
\label{eq2.13}
\alpha^{\rm LSO} = \frac{\sqrt 2}{6} \, \frac{m_1}{m_2} = \frac{1}{\sqrt{18}} \, \frac{m_1}{m_2} \, .
\end{equation}

In other words
\begin{equation}
\label{eq2.14}
G m_2 \, \Omega_{\rm phys}^{\rm LSO} = 6^{-3/2} \left[ 1 + \left( c_{\Omega}^{\rm BS} - \frac{1}{\sqrt{18}} \right) \frac{m_1}{m_2} + {\mathcal O} \left( \left( \frac{m_1}{m_2} \right)^2 \right) \right] \, .
\end{equation}

Note that the numerical correction $1/\sqrt{18} \simeq 0.235702$ leads to a quite significant decrease (by nearly 50\%) of the raw Lorenz-gauge coefficient $c_{\Omega}^{\rm BS}$, Eq.~(\ref{eq2.11}):
\begin{equation}
\label{eq2.15}
c_{\Omega}^{\rm BS} - \frac{1}{\sqrt{18}} = 0.2513(6) \, .
\end{equation}
We recommend that, in the future, the results of any Lorenz-gauge (dimensionful) quantity be explicitly reexpressed in physical units, so as to avoid the need of doing such a posteriori renormalizations. A simple way of doing so would be to explicitly change the time coordinate, \`a la Eq.~(\ref{eq2.8}), to work in a ``renormalized Lorenz gauge'' where the metric becomes asymptotically flat.

\smallskip

To compare the renormalized GSF result (\ref{eq2.14}) to the theoretical predictions of the EOB formalism, it is useful to introduce another ``renormalization''. Indeed, the natural quantity for adimensionalizing a frequency in the EOB formalism is $(GM)^{-1} \equiv (G(m_1 + m_2))^{-1}$, rather than $(Gm_2)^{-1}$, as used in GSF studies. This leads us to reexpressing (\ref{eq2.14}) as
\begin{equation}
\label{eq2.16}
GM \, \Omega_{\rm phys}^{\rm LSO} \equiv G(m_1 + m_2) \, \Omega_{\rm phys}^{\rm LSO} = 6^{-3/2} [1 + c_{\Omega}^{\rm ren} \, \nu + {\mathcal O} (\nu^2)] \, ,
\end{equation}
where the ``EOB-renormalized'' frequency sensitivity coefficient $c_{\Omega}^{\rm ren}$ is given by
\begin{equation}
\label{eq2.17}
c_{\Omega}^{\rm ren} = 1 + c_{\Omega}^{\rm BS} - \frac{1}{\sqrt{18}} = 1.2513(6) \, .
\end{equation}

As a final reexpression, it is useful to replace any adimensionalized orbital frequency $M\Omega$ by the quantity
\begin{equation}
\label{eq2.18}
x(\Omega) \equiv \left( \frac{GM \, \Omega}{c^3} \right)^{2/3} \, ,
\end{equation}
which plays a crucial role in PN and EOB developments. [Here, we have inserted the velocity of light $c$ (elsewhere set to one) as a convenient mnemonics to remember the exponent $2/3$. Indeed, $x \sim v^2 / c^2$ is the basic PN expansion parameter, and the exponent $2/3$ converts the $1/c^3$ factor into a $1/c^2$ one.] In terms of $x$, Eq.~(\ref{eq2.18}), Eqs.~(\ref{eq2.16}), (\ref{eq2.17}) become
\begin{equation}
\label{eq2.19}
x^{\rm LSO} = \frac{1}{6} \, [1 + c_x^{\rm ren} \, \nu + {\mathcal O} (\nu^2)] \, ,
\end{equation}
with
\begin{equation}
\label{eq2.20}
c_x^{\rm ren} = \frac{2}{3} \, c_{\Omega}^{\rm ren} = \frac{2}{3} \left( 1 + c_{\Omega}^{\rm BS} - \frac{1}{\sqrt{18}} \right) = 0.8342(4) \, .
\end{equation}

It is the number $c_x^{\rm ren}$, Eq.~(\ref{eq2.20}), that we shall primarily use in the following to compare the GSF results to the predictions of the EOB formalism. Let us note in passing that all the EOB estimates \cite{Buonanno:1998gg,Damour:2000we} of the mass-ratio (i.e. $\nu$) dependence of the total-mass-adimensionalized LSO frequency $GM \, \Omega^{\rm LSO} \equiv \hat\Omega^{\rm LSO} (\nu)$, as well as its other possible PN estimates \cite{Damour:2000we,Blanchet:2001id}, agree in predicting that $\hat\Omega^{\rm LSO} (\nu)$ is an {\it increasing} function of $\nu$, i.e. that the sensitivity coefficients $c_{\Omega}^{\rm ren}$, in Eq.~(\ref{eq2.16}), or $c_x^{\rm ren}$, in Eq.~(\ref{eq2.19}), are strictly {\it positive}. We just wish to remark here that this property is not necessarily linked to the Lorenz-gauge, $m_2$-adimensionalized sensitivity parameter $c_{\Omega}^{\rm BS}$, Eq.~(\ref{eq2.10}), being positive. The positivity of $c_{\Omega}^{\rm ren}$ is equivalent to the much weaker condition $c_{\Omega}^{\rm BS} > \frac{1}{\sqrt{18}} - 1 \simeq -0.76430$.

\section{A short review of the conservative EOB formalism}
\setcounter{equation}{0}

The EOB formalism comprises three main building blocks: (i) a conservative dynamics; (ii) the inclusion of radiation reaction effects; and (iii) the construction of a resummed waveform. Let us review here the item (i), which suffices for the comparison to the corresponding {\it conservative} GSF effects.

\smallskip

The conservative EOB dynamics is defined by a Hamiltonian of the form
\begin{equation}
\label{eq3.1}
H_{\rm EOB} = M \, \sqrt{1+2\nu (\hat H_{\rm eff} - 1)} \, ,
\end{equation}
where $\hat H_{\rm eff} \equiv H_{\rm eff} / \mu$ denotes the ``effective Hamiltonian'' (per unit $\mu$-mass) of an ``effective particle'' of mass $\mu$, following (modulo quartic, and possibly higher-order, terms in the radial momentum $p_r$), a geodesic in the ``effective metric''\footnote{The radial metric coefficient $g_{rr}^{\rm eff}$ is here denoted $\bar B (r)$, instead of the notation $B(r)$ used in the original EOB articles, because we wish to keep the notation $B$ for another use; see below.}
\begin{equation}
\label{eq3.2}
ds_{\rm eff}^2 = g_{\mu\nu}^{\rm eff} (x) \, dx^{\mu} dx^{\nu} = - A (r;\nu) \, dt^2 + \bar B (r,\nu) \, dr^2 + r^2 (d\theta^2 + \sin^2 \theta d\varphi^2) \, .
\end{equation}
More precisely, the conserved energy of the effective dynamics for $\mu$, $H_{\rm eff} \equiv -p_0$, is obtained by solving an ``effective one-body Hamilton-Jacobi'' equation of the form
\begin{equation}
\label{eq3.3}
0 = \mu^2 + g_{\rm eff}^{\mu\nu} (x) \, p_{\mu} \, p_{\nu} + Q(p) \, ,
\end{equation}
where $p_{\mu} = \partial S(x) / \partial x^{\mu}$ ($S$ denoting the action), and where
\begin{equation}
\label{eq3.4}
Q(p) = A^{\mu\nu\rho\sigma} (x) \, p_{\mu} \, p_{\nu} \, p_{\rho} \, p_{\sigma} + \ldots
\end{equation}
denotes quartic-in-momenta, and possibly higher-order in $p$, contributions. As shown in \cite{Damour:2000we}, one can, at the 3PN level, restrict the quartic correction (\ref{eq3.4}) to depend only on the spatial components $p_i$ of the momentum, and, more precisely, to be proportional to the fourth power of the radial momentum: $Q \propto p_r^4$. We shall assume here that the fact that $Q$ depends only on $p_r$ and $r$, and vanishes (at least) like $p_r^4$ when $p_r \to 0$ remains true at higher PN levels. Solving the Hamilton-Jacobi equation (\ref{eq3.3}) with respect to $H_{\rm eff} = -p_0 > 0$ leads to
\begin{equation}
\label{eq3.5}
H_{\rm eff} = \sqrt{A(r;\nu) \left( \mu^2 + \frac{J^2}{r^2} + \frac{p_r^2}{\bar B (r;\nu)} + Q(r,p_r)\right)}
\end{equation}
where $J$ denotes the conserved effective angular momentum. For motions in the equatorial plane $\theta = \frac{\pi}{2}$ of the coordinate system of (\ref{eq3.2}), one has simply $J = p_{\varphi}$. One of the basic principles of the EOB formalism is that the effective angular momentum $J$ is identified with the total angular momentum of the binary system. By contrast, the total energy of the binary system, ${\mathcal E} = M +$ binding energy, is not equal to the effective energy, say, ${\mathcal E}_{\rm eff} = H_{\rm eff}$, Eq.~(\ref{eq3.5}), but to the EOB Hamiltonian (\ref{eq3.1}): ${\mathcal E} = H_{\rm EOB}$. Still, we see from (\ref{eq3.5}) that $H_{\rm eff}$ governs, for a given value of $J$, the radial motion:
\begin{equation}
\label{eq3.6}
H_{\rm eff}^2 (r,p_r ; J) = A(r;\nu) \left( \mu^2 + \frac{J^2}{r^2} + \frac{p_r^2}{\bar B (r;\nu)} + Q(r,p_r) \right) \, ,
\end{equation}
where, at the 3PN level,
\begin{equation}
\label{eq3.7}
Q(r,p_r) = 2 \, (4-3\nu) \, \nu \, \frac{(GM)^2}{r^2} \, \frac{p_r^4}{\mu^2} \, .
\end{equation}

In practical calculations, it is quite useful to rescale the various quantities, $r , p_r , J , {\mathcal E}_{\rm eff}$ into dimensionless ones, using
\begin{equation}
\label{eq3.8}
t = GM \, \hat t \, , \quad r \equiv GM \, \hat r \, , \quad p_r \equiv \mu \, \hat p_r \, , \quad J \equiv GM \, \mu \, j \, , \quad {\mathcal E}_{\rm eff} \equiv \mu \, \hat{\mathcal E}_{\rm eff} \, .
\end{equation}
Note that the effective action
\begin{equation}
\label{eq3.9}
S = - {\mathcal E}_{\rm eff} \, t + J \, \varphi + S_{\rm rad} (r) \, ,
\end{equation}
is rescaled into $\hat S \equiv S/GM \, \mu$, so that, for instance, $\hat p_r$ is canonically conjugate to $\hat r$ with respect to the (rescaled) symplectic structure defined by $\hat S$. The rescaled version of Eqs.~(\ref{eq3.6}), (\ref{eq3.7}) then reads (at the 3PN level)
\begin{equation}
\label{eq3.10}
\hat H_{\rm eff}^2 (\hat r , \hat p_r ; j) = A(\hat r ; \nu) \left( 1 + \frac{j^2}{\hat r^2} + \frac{\hat p_r^2}{\bar B (\hat r ; \nu)} + \hat Q (\hat r , \hat p_r)\right) \, ,
\end{equation}
where, at the 3PN level,
\begin{equation}
\label{eq3.11}
\hat Q (\hat r , \hat p_r) = 2 \, (4-3\nu) \, \nu \, \frac{\hat p_r^4}{\hat r^2} \, .
\end{equation}

In the present work we shall, like Ref.~\cite{Barack:2009ey}, mostly focus on small-eccentricity orbits, i.e. on the limits where $\hat p_r^2 \ll j^2 / \hat r^2$. Then, working to first-order in the squared eccentricity $e^2 \sim \hat p_r^2$ we can neglect the quartic (or more) contribution $\hat Q$ which is ${\mathcal O} (e^4)$. We then see that, in this case, the conservative EOB dynamics depends on the knowledge of two functions of $\hat r$ and $\nu$, namely the metric functions $A(\hat r ; \nu)$ and $\bar B (\hat r ; \nu)$ entering the effective metric (\ref{eq3.2}). In previous EOB works, it was found convenient to replace the second metric function $\bar B = g_{rr}^{\rm eff}$ by the combination
\begin{equation}
\label{eq3.12}
D(r) \equiv A(r) \, \bar B(r) \, .
\end{equation}
Here, we shall find it even more convenient to use instead the {\it inverse} of the $D$ function (which naturally entered Ref.~\cite{Damour:2001tu}, where it was proposed to resum $D$ by working with the Taylor expansion of its inverse). We shall denote it as
\begin{equation}
\label{eq3.13}
\bar D (r) \equiv D^{-1} (r) \equiv (A(r) \, \bar B(r))^{-1} \, .
\end{equation}
Another useful EOB notation is to introduce a special notation for the inverse of the scaled EOB radius $\hat r$, namely
\begin{equation}
\label{eq3.14}
u \equiv \frac{1}{\hat r} \equiv \frac{GM}{r} \, .
\end{equation}
With this notation in hand, the conservative EOB dynamics of small-eccentricity orbits depends on the knowledge of two functions: $A(u;\nu)$ and $\bar D (u;\nu)$. At the 3PN level, these two functions have been found to take the values \cite{Damour:2000we}
\begin{equation}
\label{eq3.15a}
A (u;\nu) = 1-2u + 2\nu \, u^3 + \nu \, a_4 \, u^4 + {\mathcal O} (u^5) \, ,
\end{equation}
\begin{equation}
\label{eq3.15b}
\bar D (u;\nu) = 1+6\nu \, u^2 + 2(26-3\nu) \, \nu \, u^3 + {\mathcal O} (u^4) \, ,
\end{equation}
where $a_4 = 94/3 - 41 \pi^2 / 32$.

\smallskip

As summarized in Eqs.~(\ref{eq3.15a}), (\ref{eq3.15b}) the PN calculations give access to the first few terms of the expansions of $A$ and $\bar D$ {\it in powers of} $u$, for a fixed value of the symmetric mass ratio $\nu$. If we formally extend the definition of the EOB formalism, and of the $A$ and $\bar D$ functions, to an arbitrary PN order (as was shown to be possible in Section~III of \cite{Damour:2000we}), assuming that the ``correction'' term $Q$ is kept of order $p_r^4$ as $p_r \to 0$, we can think of $A(u;\nu)$ and $\bar D (u;\nu)$ as being some yet unknown mathematical functions of {\it two} variables.

\smallskip

At present, one has in hands two different computational tools for acquiring some knowledge of the unknown ``abstract'' EOB potentials $A(u;\nu)$ and $\bar D (u;\nu)$. On the one hand, existing PN calculations (at the 3PN level) have given us access to the $u$-expansions (\ref{eq3.15a}), (\ref{eq3.15b}) (and to the first term (\ref{eq3.11}) in the $Q$ term). On the other hand, NR simulations gives us access (when $\nu$ is not too different from $1/4$) to various data that depend on $A$ and $\bar D$ and can therefore be used, in principle, to map the two-dimensional profiles of $A(u;\nu)$ and $\bar D (u;\nu)$. The first attempt to do so dates from 2002. Ref.~\cite{Damour:2002gh} compared NR data on the gravitational binding energy of (corotating and waveless\footnote{i.e. admitting a helical Killing vector.}) circular binary systems to the EOB predictions, and looked for the best fit to a {\it generalized} $A$ function (\ref{eq3.15a}), extended by an additional (4PN) term $+ \, a_5 (\nu) \, u^5$. [Note in passing that their result (for $\nu = 1/4$) suggested a negative value for the coefficient of $u^5$, $a_5 ( \mbox{$\frac{1}{4}$} ) \simeq -3$, i.e. a slope $a_5 \simeq -12$ if one considers a linear dependence on $\nu : a_5 (\nu) = a_5 \, \nu$. Such a negative value is not unreasonable, as we shall further discuss below.] More recently, many works \cite{Buonanno:2007pf,Damour:2007yf,Damour:2007vq,Damour:2008te,Boyle:2008ge,Damour:2009kr,Buonanno:2009qa} have compared NR data on the waveform emitted by radiation-reaction-driven, inspiralling binary black holes to the predictions of several {\it extended} versions of the EOB formalism. These extensions consisted in adding parametrized extra contributions not only to the PN expansion of the $A$ function ($\delta A^{\rm PN} = \nu \, a_5 \, u^5 + \nu \, a_6 \, u^6$), but also to the other building blocks of the EOB formalism: the radiation reaction force, and the emitted waveform. We shall compare below the results of the most recent NR/EOB comparison of this type, and the recent GSF result (\ref{eq2.19}), (\ref{eq2.20}).

\smallskip

The main purpose of the present work is to explicate how GSF results can also be used as a third computational tool for acquiring some knowledge about the conservative EOB potentials $A(u;\nu)$ and $\bar D (u;\nu)$. The first point to emphasize is that current GSF technology can, at most, give us access to the first terms in the expansions of the functions $A(u;\nu)$ and $\bar D (u;\nu)$ in powers of the symmetric mass ratio $\nu$. More precisely, let us consider the expansions of the (unknown) exact functions $A(u;\nu)$ and $\bar D (u;\nu)$ in powers of $\nu$, say
\begin{equation}
\label{eq3.16a}
A(u;\nu) = 1-2u + \nu \, a (u) + \nu^2 \, a_2 (u) + {\mathcal O} (\nu^3) \, ,
\end{equation}
\begin{equation}
\label{eq3.16b}
\bar D(u;\nu) = 1 + \nu \, \bar d (u) + \nu^2 \, \bar d_2 (u) + {\mathcal O} (\nu^3) \, ,
\end{equation}
where, for notational simplicity, we have suppressed the index 1 on the contributions which are linear in $\nu$: $\nu \, a_1 (u) \equiv \nu \, a(u)$, $\nu \, \bar d_1 (u) \equiv \nu \, \bar d (u)$. In the following, we shall show to what extent GSF studies can inform us about the two $\nu$-linear functions of $u$, $a(u)$ and $\bar d (u)$.

\smallskip

Before plunging into the details of how GSF studies can tell us something about the two functions $a(u)$ and $\bar d (u)$ entering the conservative EOB dynamics\footnote{To which should be added another function (of $u$ and $p_r$) parametrizing the linear-in-$\nu$ piece in the the ``exact'' version of the $\hat Q$ contribution in Eq.~(\ref{eq3.10}). As said above, we shall consider here observables that do not depend on the $Q$ contribution.}, let us emphasize the complementarity of the expansions (\ref{eq3.15a}), (\ref{eq3.15b}) versus the expansions (\ref{eq3.16a}), (\ref{eq3.16b}). The PN expansions (\ref{eq3.15a}), (\ref{eq3.15b}) proceed in powers of $u = GM/c^2 r$, but compute, at a given order in $u$, the exact dependence on $\nu$. On the other hand, the ``GSF'' expansions (\ref{eq3.16a}), (\ref{eq3.16b}) proceed in powers of $\nu$, and can in principle give access, at a given order in $\nu$, to the exact dependence on $u$. For instance, the first-order GSF coefficients $a(u)$ and $\bar d (u)$, if known exactly, would give us information about some arbitrarily high PN contributions in (\ref{eq3.15a}), (\ref{eq3.15b}). Reciprocally, as the PN results (\ref{eq3.15a}), (\ref{eq3.15b}) have made no truncation on the powers of $\nu$, we see that Eqs.~(\ref{eq3.15a}), (\ref{eq3.15b}) is already giving us information about high-order terms in the ``GSF expansions'' (\ref{eq3.16a}), (\ref{eq3.16b}) in powers of $\nu$. In particular, the (remarkable) fact that the (vanishing) 1PN, 2PN $(a_3)$ and 3PN $(a_4)$ contributions to the $A$ potential are linear in $\nu$ is already telling us something about both the ``second-order GSF'' (2 GSF) contribution $\nu^2 \, a_2 (u)$, and the ``third-order GSF'' (3 GSF) contribution $\nu^3 \, a_3 (u)$. Indeed, at the $n$PN order (which corresponds to a term $\propto u^{n+1}$ in $A(u)$) one finds that, in intermediate calculations, the coefficient of $u^{n+1}$ is a polynomial in $\nu$ of degree $n$. It was, however, found in Ref.~\cite{Damour:2000we} that remarkable cancellations take place in the computation of the $A$ function, and that the $\nu^2$ terms present at 2PN $(\propto u^3)$, and {\it both} the $\nu^2$ terms and the $\nu^3$ terms present at 3PN $(\propto u^4)$ exactly cancell in the final result for $A(u;\nu)$. From a practical point of view, GSF studies have not yet embarked on any real ``second-order GSF'' work. We shall therefore focus, in the following, on the ``first-order GSF'' contributions $\nu \, a(u)$ and $\nu \, \bar d (u)$ in (\ref{eq3.16a}), (\ref{eq3.16b}).

\section{Circular orbits in the EOB formalism and the Last Stable (circular) Orbit (LSO)}
\setcounter{equation}{0}

The stable circular orbits in the EOB formalism are conveniently discussed by using the dimensionless, rescaled variables (\ref{eq3.8}) and by considering the squared effective Hamiltonian for the radial motion, Eq.~(\ref{eq3.10}). As usual the presence of a positive kinetic energy term, $\hat p_r^2 / \bar B + \hat Q (\hat p_r)$, associated to the radial momentum, on the right-hand-side (RHS) of (\ref{eq3.10}) implies that the stable circular orbits (for a given dimensionless angular momentum $j$) correspond to {\it minima} (with respect to $\hat r$) of the ``effective radial potential''
\begin{equation}
\label{eq4.1}
A(\hat r ; \nu) \left( 1+\frac{j^2}{\hat r^2} \right) \, .
\end{equation}

Using the short-hand notation (\ref{eq3.14}), we shall then define the function
\begin{equation}
\label{eq4.2}
W_j (u;\nu) \equiv A(u;\nu)(1+j^2 u^2) \equiv A(u;\nu) + j^2 B(u;\nu) \, ,
\end{equation}
where we also introduced the short-hand
\begin{equation}
\label{eq4.3}
B(u;\nu) \equiv u^2 A(u;\nu)
\end{equation}
which should not be confused with the metric component $\bar B (u;\nu) \equiv g_{rr}^{\rm eff}$ entering (\ref{eq3.2}).

\smallskip

Stable circular orbits correspond to minima (with respect  to $u$) of $W_j (u)$, i.e. they solve $W'_j (u) = 0$, with $W''_j (u) > 0$. [Here, and in the following, a prime will denote a $u$-derivative.] The solutions of $W'_j (u) = 0$ with $W''_j (u) < 0$ correspond to unstable circular orbits, while the LSO is the solution of $W'_j (u) = 0$ which satisfies $W''_j (u) = 0$.

\smallskip

We can then parametrize the (one-parameter) sequence of circular orbits by the value of $u = 1/\hat r$. Indeed, while it is a priori difficult to solve (suppressing the presence of $\nu$ in $A$ and $B$)
\begin{equation}
\label{eq4.4}
W'_j (u) = A'(u) + j^2 B'(u) = 0 \, ,
\end{equation}
with respect to $u$, it is trivial to solve it with respect to $j^2$, namely
\begin{equation}
\label{eq4.5}
j_{\rm circ}^2 (u) \equiv - \frac{A'(u)}{B'(u)} \, .
\end{equation}

Knowing $u$ and $j_{\rm circ} (u)$ one can then compute all the physical quantities attached to the circular orbit. From (\ref{eq3.10}) (and $p_r = 0$) the (specific) effective energy is
\begin{equation}
\label{eq4.6}
\hat H_{\rm eff}^{\rm circ} (u) = \sqrt{A(u) + j_{\rm circ}^2 (u) \, B(u)} \, ,
\end{equation} 
while the corresponding total energy (divided by the total mass $M$) is
\begin{equation}
\label{eq4.7}
h(u) \equiv \frac{H_{\rm EOB}^{\rm circ}}{M} = \sqrt{1+2\nu (\hat H_{\rm eff}^{\rm circ} (u) - 1)} \, .
\end{equation}

Finally the orbital frequency around the circular orbit is obtained from Hamilton's equation of motion for $\varphi$:
\begin{equation}
\label{eq4.8}
\Omega \equiv \frac{d\varphi}{dt} = \frac{\partial \, H_{\rm EOB} (r,p_r,p_{\varphi})}{\partial \, p_{\varphi}} = \frac{\partial \, H_{\rm EOB}}{\partial \, J} \, .
\end{equation}

This yields
\begin{equation}
\label{eq4.9}
\hat\Omega^{\rm circ} (u) \equiv GM \, \Omega^{\rm circ} (u) = \frac{j_{\rm circ} (u) \, B(u)}{h(u) \, \hat H_{\rm eff} (u)} \, .
\end{equation}

Squaring (\ref{eq4.9}), and inserting (\ref{eq4.5})--(\ref{eq4.7}) in the result (remembering that $B(u) \equiv u^2 A(u)$), finally leads to the following simple result
\begin{equation}
\label{eq4.10}
\frac{\hat\Omega_{\rm circ}^2 (u)}{u^3} = - \frac{1}{2} \, \frac{A' (u)}{h^2 (u)} = - \frac{1}{2} \, \frac{A'(u)}{1+2\nu (\hat H_{\rm eff}^{\rm circ} (u)-1)} \, .
\end{equation}
The result (\ref{eq4.10}) is equivalent to Eq.~(11) in \cite{Damour:2008te}. Replacing $\hat\Omega = GM \, \Omega$ by the corresponding $x(\Omega) \equiv (GM \, \Omega)^{2/3}$, Eq.~(\ref{eq2.18}), we can rewrite (\ref{eq4.10}) as
\begin{equation}
\label{eq4.11}
x_{\rm circ} (u) = u \left( \frac{-\frac{1}{2} \, A'(u)}{1+2\nu (\hat H_{\rm eff}^{\rm circ} (u) - 1)} \right)^{1/3} \, ,
\end{equation}
where the explicit expression of $\hat H_{\rm eff}^{\rm circ} (u)$ would be (from (\ref{eq4.5}) and (\ref{eq4.6}))
\begin{equation}
\label{eq4.12}
\hat H_{\rm eff}^{\rm circ} (u) = \sqrt{A(u) - \frac{A'(u)}{B'(u)} \, B(u)} = \sqrt{\frac{2u \, A^2 (u)}{2u \, A(u) + u^2 A'(u)}} \, .
\end{equation}
Up to now we have made no approximations. In particular, we can conclude that the exact LSO frequency parameter $x_{\rm LSO} = (GM \, \Omega_{\rm LSO})^{2/3}$ is obtained by inserting in (\ref{eq4.11}) the root $u_{\rm LSO}$ of
\begin{equation}
\label{eq4.13}
0 = \Delta (u) \equiv A'(u) \, B''(u) - A''(u) \, B'(u) \, ,
\end{equation}
which is the condition for having a common solution to $W'_j = A' + j^2 B' = 0$ and $W''_j = A'' + j^2 B'' = 0$. A more explicit form of the discriminant $\Delta$ is
\begin{equation}
\label{eq4.14}
\Delta = 2 \, A \, A' + 4 \, u \, (A')^2 - 2 \, u \, A \, A'' \, .
\end{equation}
As a simple check on the above results, and as a warm up for later, one easily sees that, in the test-mass limit $\nu \to 0$, i.e. when $A(u) \to 1 - 2u$, one recovers well-known results for a test mass in circular orbit around a Schwarzschild black hole\footnote{We indeed recall that the EOB formalism is constructed so that the $\nu \to 0$ limit yields the dynamics of a test mass in a Schwarzschild spacetime.}
\begin{equation}
\label{eq4.15}
j_{\rm circ} (u) = \frac{1}{\sqrt{u(1-3u)}} + {\mathcal O} (\nu) \, ,
\end{equation}
\begin{equation}
\label{eq4.16}
\hat H_{\rm eff}^{\rm circ} (u) = \frac{1-2u}{\sqrt{1-3u}} + {\mathcal O} (\nu) \, ,
\end{equation}
\begin{equation}
\label{eq4.17}
(GM \, \Omega_{\rm circ} (u))^{2/3} = x_{\rm circ} (u) = u +  {\mathcal O} (\nu) \, .
\end{equation}
As for the LSO condition (\ref{eq4.13}) it reduces in this limit to
\begin{equation}
\label{eq4.18}
0 = \Delta (u) = 4 (6u-1) + {\mathcal O} (\nu) \, ,
\end{equation}
so that one recovers the well-known $u_{\rm LSO} = GM/r_{\rm LSO} = \frac{1}{6} + {\mathcal O} (\nu)$.

\smallskip

One should also note that the full dynamics of circular orbits is entirely described by a single function, namely the EOB ($-g_{00}^{\rm eff}$) metric component $A(u;\nu)$. In particular the value of the LSO frequency parameter $x_{\rm LSO} (\nu)$ can only depend on $A(u;\nu)$.

\smallskip

Let us now insert in the general results (\ref{eq4.5})--(\ref{eq4.12}) the ``GSF expansion'' of the EOB $A$ potential, i.e. the expansion (\ref{eq3.16a}) of the function $A(u;\nu)$ in powers of the symmetric mass ratio $\nu$. Keeping only the first-order correction in $\nu$ (``1 GSF approximation''), the GSF-expansion of Eq.~(\ref{eq4.5}) yields
\begin{equation}
\label{eq4.19}
j_{\rm circ} (u) = \frac{1}{\sqrt{u(1-3u)}} \left( 1- \frac{1}{4} \, \nu \, a'(u) - \frac{1}{4} \, \nu \, \frac{b'(u)}{u(1-3u)} + {\mathcal O} (\nu^2) \right) \, ,
\end{equation}
where we introduced the short-hand notation
\begin{equation}
\label{eq4.20}
b(u) \equiv u^2 \, a(u) \, ,
\end{equation}
so that $B(u;\nu) = u^2 (1-2u) + \nu \, b(u) + {\mathcal O} (\nu^2)$. Concerning the GSF-expansion of $\hat H_{\rm eff}$, the zeroth-order (test-mass) result (\ref{eq4.16}) is sufficient for our present purpose because $\hat H_{\rm eff}$ enters the frequency parameter $x$, Eq.~(\ref{eq4.11}), multiplied by a factor $\nu$. The GSF-expansion of Eq.~(\ref{eq4.11}) yields
\begin{equation}
\label{eq4.21}
x = u \left( 1-\frac{1}{6} \, \nu \, a'(u) - \frac{2}{3} \, \nu \left( \frac{1-2u}{\sqrt{1-3u}} - 1 \right) + {\mathcal O} (\nu^2) \right) \, ,
\end{equation}
or, reciprocally,
\begin{equation}
\label{eq4.22}
u = x \left( 1+\frac{1}{6} \, \nu \, a'(x) + \frac{2}{3} \, \nu \left( \frac{1-2x}{\sqrt{1-3x}} - 1 \right) + {\mathcal O} (\nu^2) \right) \, .
\end{equation}

To determine the location of the LSO we also need the GSF expansion of the discriminant $\Delta$, Eq.~(\ref{eq4.13}). The expansion of Eq.~(\ref{eq4.14}) yields
\begin{equation}
\label{eq4.23}
- \frac{1}{4} \, \Delta (u) = 1 - 6u + \nu \, \bar a (u) + {\mathcal O} (\nu^2) \, ,
\end{equation}
where we introduced the short-hand
\begin{equation}
\label{eq4.24}
\bar a(u) \equiv a(u) + \frac{1}{2} (10 u - 1) \, a'(u) + \frac{1}{2} \, u (1-2u) \, a'' (u) \, .
\end{equation}

Solving the LSO condition $\Delta (u) = 0$ then immediately leads to
\begin{equation}
\label{eq4.24bis}
u_{\rm LSO} = \frac{1}{6} \left[ 1+ \nu \, \bar a ( \mbox{$\frac{1}{6}$} ) + {\mathcal O} (\nu^2) \right] \, ,
\end{equation}
where
\begin{equation}
\label{eq4.25}
\bar a ( \mbox{$\frac{1}{6}$} ) = a ( \mbox{$\frac{1}{6}$} ) + \frac{1}{3} \, a' ( \mbox{$\frac{1}{6}$} ) + \frac{1}{18} \, a'' ( \mbox{$\frac{1}{6}$} ) \, .
\end{equation}

Inserting the result (\ref{eq4.24}) in Eq.~(\ref{eq4.21}) finally yields the EOB prediction for the frequency parameter of the LSO to first order in $\nu$,
\begin{equation}
\label{eq4.26}
x_{\rm LSO} = \frac{1}{6} \, [1 + c_x^{\rm EOB} \, \nu + {\mathcal O} (\nu^2)]
\end{equation}
with
\begin{equation}
\label{eq4.27n}
c_x^{\rm EOB} = c_x^E + \tilde a ( \mbox{$\frac{1}{6}$} ) \, ,
\end{equation}
where
\begin{equation}
\label{eq4.27}
c_x^E \equiv \frac{2}{3} \left( 1-\sqrt{\frac{8}{9}} \right) \, ,
\end{equation}
and
\begin{equation}
\label{eq4.28}
\tilde a ( \mbox{$\frac{1}{6}$} ) \equiv a ( \mbox{$\frac{1}{6}$} ) + \frac{1}{6} \, a' ( \mbox{$\frac{1}{6}$} ) + \frac{1}{18} \, a'' ( \mbox{$\frac{1}{6}$} ) \, .
\end{equation}
[Note that $\tilde a (1/6)$ differs from $\bar a (1/6)$, Eq.~(\ref{eq4.25}), in the coefficient of $a'$.] The parameter $c_x^{\rm EOB}$ measures the fractional sensitivity of $x_{\rm LSO}$ to $\nu$, around $\nu = 0$. It is also related to the fractional $\nu$-sensitivity $c_{\Omega}^{\rm EOB}$ of the $M$ adimensionalized LSO orbital frequency $GM \, \Omega_{\rm LSO}$,
\begin{equation}
\label{eq4.29}
GM \, \Omega_{LSO} \equiv (x_{\rm LSO})^{3/2} = \frac{1}{6^{3/2}} \, [1+c_{\Omega}^{\rm EOB} \, \nu + {\mathcal O} (\nu^2)] \, ,
\end{equation}
by
\begin{equation}
\label{eq4.30}
c_{\Omega}^{\rm EOB} = \frac{3}{2} \, c_x^{\rm EOB} \, .
\end{equation}
As we see from (\ref{eq4.27n}), $c_x^{\rm EOB}$ is made of two separate contributions: (i) a numerical contribution which is independent of the function $a(u)$ (and which comes from the specific EOB ``energy map'' (\ref{eq3.1}) relating the ``effective'' energy to the ``real'' energy); and (ii) a contribution which is a linear combination of the values at the unperturbed LSO of $a(u)$ and its first two derivatives.

\smallskip

Let us now compare the EOB prediction (\ref{eq4.27n}) to the recent GSF result (\ref{eq2.20}). We shall consider here the case where $a(u)$ is given by a simple power-law PN expansion 
\begin{equation}
\label{eq4.31}
a(u) = \sum_{n \geq 3} a_n \, u^n = a_3 \, u^3 + a_4 \, u^4 + a_5 \, u^5 + a_6 \, u^6 + a_7 \, u^7 + \ldots
\end{equation}
which is the type currently assumed in the NR-EOB comparisons. [Below, we shall argue that the coefficients $a_5 , a_6 , \ldots$ must include a logarithmic dependence on $u$.] Inserting the expansion (\ref{eq4.31}) in Eq.~(\ref{eq4.28}), we get (if we neglect any ``logarithmic running'' of $a_5 , a_6 , \ldots$)
\begin{equation}
\label{eq4.32}
\tilde a ( \mbox{$\frac{1}{6}$} ) = \sum_{n \geq 3} \tilde a_n = \tilde a_3 + \tilde a_4 + \tilde a_5 + \tilde a_6 + \tilde a_7 + \ldots
\end{equation}
where
\begin{equation}
\label{eq4.33}
\tilde a_n \equiv (2 \, n^2  - n +1) \, \frac{a_n}{6^n} \, .
\end{equation}
The currently known coefficients $a_n$ in the PN expanqion of $A(u;\nu)$ \cite{Damour:2000we} (see Eq.~(\ref{eq3.15a})) are
\begin{eqnarray}
\label{eq4.34}
a_3 &= &2 \, , \nonumber \\
a_4 &= &\frac{94}{3} - \frac{41 \, \pi^2}{32} = 18.687903 \, . 
\end{eqnarray}

The contributions of the currently known terms in the PN expansion of $A$ are, numerically,
\begin{eqnarray}
\label{eq4.35}
c_x^E &= &0.038127 \, , \\
\label{eq4.36}
\tilde a_3 &= &\frac{4}{27} = 0.148148 \, , \\
\label{eq4.37}
\tilde a_4 &= &\frac{29}{1296} \, a_4 = 0.418171 \, ,
\end{eqnarray}
so that they add up to
\begin{equation}
\label{eq4.38}
c_x^E + \tilde a_3 + \tilde a_4 = 0.604446 \, .
\end{equation}
Comparing this (partial) result to the GSF estimate (\ref{eq2.20}), we see that the 3PN approximation to $a(u)$ explains about 72.5\% of the GSF result. We expect that the missing 27.5\% will be contributed by the higher PN contributions to $a(u)$, coming from $a_5 , a_6 , a_7 , \ldots$. More precisely, by considering the difference between (\ref{eq4.32}) and (\ref{eq2.20}), we should expect (when neglecting any logarithmic running) that
\begin{equation}
\label{eq4.40}
\frac{23}{3888} \, a_5 + \frac{67}{46656} \, a_6 + \frac{23}{69984} \, a_7 + \ldots = 0.22975(40) \, .
\end{equation}
Eq.~(\ref{eq4.40}) encapsulates the knowledge about higher-order (4PN, 5PN, 6PN,$\ldots$) contributions to the function $a(u) \equiv [\partial \, A (u;\nu)/\partial \, \nu]_{\nu = 0}$ contained in the recent GSF work \cite{Barack:2009ey}. We can roughly estimate the order of magnitude of the successive terms on the left-hand side (L.H.S.) of (\ref{eq4.40}) by evaluating the (inverse) radius of convergence $\rho$ of the series (\ref{eq4.31}) from the $a_4$-coefficient, say $\rho \simeq \rho_4 \equiv a_4^{1/4} = 2.079171$. As $\rho_4$ is quite close to $2$, it looks reasonable to expect that $a_n$ will roughly grow as $\varepsilon_n \, 2^n$ where $\varepsilon_n = \pm$ is a sign. This would mean that the higher-order contributions $\tilde a_n$ to $c_x^{\rm EOB}$ decrease as $\varepsilon_n (2 \, n^2 - n + 1) \, 3^{-n}$. For $n=6$, this yields $\pm \, 0.092$ which is about 11\% of the total $x_x^{\rm ren}$, Eq.~(\ref{eq2.20}), while for $n=7$ this yields $\pm \, 0.042$, which is about 5\% of $c_x^{\rm ren}$. Though the decrease with $n$ is rather slow, we can hope that the result (\ref{eq4.40}) could give us, within a rough approximation, a constraint involving mainly $a_5$ and $a_6$. Numerically this constraint would then read
\begin{equation}
\label{eq4.41}
a_5 + 0.242754 \, a_6 \simeq 38.84(7) \, ,
\end{equation}
or equivalently
\begin{equation}
\label{eq4.42}
4.11940 \, a_5 + a_6 \simeq 160.0(3) \, .
\end{equation}
The most accurate  current EOB models use an $A(u;\nu)$ potential which contains only two free parameters, denoted $a_5$ and $a_6$, and which has the following properties: (i) $A^{\rm EOB} (u,\nu,a_5,a_6)$ is a certain Pad\'e approximant, namely the ratio of two polynomials in $u$ of the form, $A^{\rm EOB} (u;\nu) = (1+ n_1 \, u) / (1+d_1 \, u + d_2 \, u^2 + d_3 \, u^3 + d_4 \, u^4 + d_5 \, u^5)$, where the coefficients $n_1 , d_1 , \ldots , d_5$ are rational functions of $\nu$, and (ii) the expansion {\it in powers of} $\nu$ of $A(u;\nu)$ is of the form (\ref{eq3.16a}) with
\begin{eqnarray}
\label{eq4.43}
a^{\rm EOB} (u) &= &a_3 \, u^3 + a_4 \, u^4 + a_5 \, u^5 + a_6 \, u^6 \, , \\
\label{eq4.44}
a_2^{\rm EOB} (u) &= &b_7 \, u^7 + b_8 \, u^8 + b_9 \, u^9 + b_{10} \, u^{10} + b_{11} \, u^{11}  \, , 
\end{eqnarray}
where the coefficients $b_7 , \ldots , b_{11}$ entering $a_2^{\rm EOB} (u)$ are polynomials in $a_3 , a_4 , a_5$, $a_6$.

\smallskip

We see on Eq.~(\ref{eq4.13}) that the $\nu$-linearized $a(u)$ function associated to these EOB models is of the general form (\ref{eq4.31}) written above, but with the restriction that $a(u)$ is a polynomial in $u$ which does not contain powers of $u$ beyond $u^6$. In addition, our notation has been chosen to be consistent in that the parameters $a_5$ and $a_6$ entering (\ref{eq4.43}) can be identified with the corresponding parameters in (\ref{eq4.31}). [In both cases too $a_3$ and $a_4$ are defined by Eq.~(\ref{eq4.34}).] Therefore, when considering the current EOB models, the GSF-LSO constraint (\ref{eq4.40}) does reduce to the linear constraint on $a_5$ and $a_6$ written in (\ref{eq4.41}) or (\ref{eq4.42}). In addition, the work of \cite{Damour:2009kr,Damour:2009ic} has shown that the constraint of having a good agreement between NR data and EOB predictions selects a long and thin (banana-shaped) region in the $a_5 , a_6$ plane. It was found by Damour and Nagar that this thin ``good fit'' region is, to a very good approximation, located around a line in the $a_5 , a_6$ plane along which $A'_{\rm EOB} \left(u_0 ; \frac{1}{4} ; a_5 , a_6\right)$ is constant (and equal, say, to its value at $(a_5 , a_6) = (-4 , 24)$ which is one of the good-fit points, lying near the leftmost part of the good-fit region). Here the prime denotes as above a $u$ derivative; the first argument $u_0 = (0.1)^{2/3} = 0.215443$ is a value of the $u$ parameter which approximately corresponds to the EOB-predicted (adiabatic) LSO frequency for the equal-mass case namely $M \, \Omega^{\rm LSO} (\nu = 1/4) \simeq 0.1$; while the second argument in $A'_{\rm EOB}$ is the value of $\nu$ corresponding to the equal-mass case $(\nu = 1/4)$, which is the case where very accurate NR data \cite{Boyle:2007ft} were used to ``tune'' the considered EOB model. By numerically constructing the analytically defined central good-fit line just defined, one finds that it intersects the GSF-LSO straight line (\ref{eq4.41}) into a unique point located at
\begin{equation}
\label{eq4.45}
a_5^{\cap} \simeq -22.3 \, , \qquad a_6^{\cap} \simeq +252 \, .
\end{equation}

These ``intersection'' ($\cap$) values for the parameters $(a_5 , a_6)$ happen to lie roughly in the middle of the banana-shaped good-fit region found in \cite{Damour:2009kr,Damour:2009ic} (indeed, the latter region extends from $(a_5 , a_6) = (0 , -20)$ to $(a_5 , a_6) = (-36 ,$ $+520)$). This suggests that the information coming from the $\nu \ll 1$ GSF-LSO study is able to break the degeneracy among $a_5$ and $a_6$ left after fitting EOB to equal-mass $\left( \nu = \frac{1}{4} \right)$ NR data.

\smallskip

At this stage, the result (\ref{eq4.45}) has only an indicative value. It is a hint that the ``real'' values of the coefficients of the (4PN and 5PN) $\nu \, u^5$ and $\nu \, u^6$ contributions to the EOB potential $A(u;\nu)$ lie near the ``intersection'' values (\ref{eq4.45}). To confirm (or refute) this hint one will need several types of further studies. First, it will be important to firm up the GSF result (\ref{eq2.20}). [In the next Section, we shall indicate ways of doing so.] Second, one will need to explore in more detail the comparison between comparable-mass NR waveforms and EOB predictions to see: (i) whether the vicinity of the values (\ref{eq4.45}) does indeed leads to a better agreement; (ii) to what extent the possible ``logarithmic running'' of $a_5$ and $a_6$ affects the estimates (\ref{eq4.45}) (see below); (iii) whether the influence of higher-PN coefficients $a_7 , \ldots$ is indeed small; and (iv) whether the current Pad\'e-based definition of the EOB potential $A(u;\nu)$ is a sufficiently accurate representation of the $A$ potential for all values of $\nu$ in the interval $0 \leq \nu \leq \frac{1}{4}$. Indeed, for the moment, the most accurate tests of the Pad\'e-constructed $A$ potential have been obtained for the equal-mass case, $\nu = 1/4$, for which the NR data are the most accurate \cite{Boyle:2007ft}. Though the NR/EOB comparisons for several unequal-mass cases have also shown a good agreement, the numerical accuracy of these tests is smaller than the one reached in the equal-mass case. For all those reasons, it does not make sense, at this stage, to indicate ``error bars'' around the values (\ref{eq4.45}).

\section{Small-eccentricity orbits in the EOB formalism}
\setcounter{equation}{0}

In the previous Section we considered {\it exactly circular} orbits in the EOB formalism and showed that the GSF determination of the $\nu$-derivative (at the test-mass limit $\nu = 0$) of the LSO orbital frequency was giving us access to a combination of radial derivatives (evaluated at the Schwarzschild LSO $u_{\rm LSO}^{\rm Schw} = GM/r_{\rm LSO}^{\rm Schw} = 1/6$) of the $\nu$-derivative $a(u) = [\partial \, A (u;\nu) / \partial \, \nu]_{\nu = 0}$ of the basic EOB radial function $A(u;\nu) \equiv -g_{00}^{\rm eff}$. Here, we shall consider (within the EOB formalism) {\it slightly non circular} orbits and show that a comparison of their gauge-invariant characteristics to GSF data can give us access to much more detailed information about the EOB formalism: in principle, one should be able to measure a certain gauge-invariant {\it function} $\tilde\rho (x)$, where $\tilde\rho (x)$ is a certain ($x$-dependent) linear combination of the $\nu$-derivatives of the {\it two} basic EOB radial functions, $A(u;\nu) = -g_{00}^{\rm eff}$ and $\bar B (u;\nu) = +g_{rr}^{\rm eff}$, entering the effective metric (\ref{eq3.2}).

\smallskip

As we shall focus, in this Section, on the ``small-eccentricity limit'' where, in the effective Hamiltonian (\ref{eq3.5}), the radial kinetic energy $p_r^2 / \bar B (r)$ (which is proportional to the square of the eccentricity $e$) is much smaller than the ``azimuthal'' kinetic energy $J^2/r^2$, we are allowed to neglect the higher-order contribution $Q(r,p_r)$ in (\ref{eq3.5}), because it is (at least) {\it quartic} in $p_r$, and therefore of higher-order $({\mathcal O} (e^4))$ in the eccentricity. [This result holds whatever be one's definition of the eccentricity $e$ in this relativistic context.] Neglecting $Q$ in the original EOB Hamilton-Jacobi equation (\ref{eq3.3}) leads to the usual Hamilton-Jacobi equation $(\mu^2 + g_{\rm eff}^{\mu\nu} \, p_{\mu} \, p_{\nu} = 0)$ describing geodesic motion in the EOB effective metric (\ref{eq3.2}). It is then convenient to describe this geodesic motion within a Lagrangian formalism, rather than the usual EOB Hamiltonian one. Using (Stueckelberg's) proper time formalism, we can then use as starting point the quadratic action
\begin{equation}
\label{eq5.1}
S = \int d\tau_{\rm eff} \, {\mathcal L} \left( x^{\mu} (\tau_{\rm eff}) , \frac{dx^{\mu} (\tau_{\rm eff})}{d\tau_{\rm eff}} \right) 
\end{equation}
where
\begin{equation}
\label{eq5.2}
{\mathcal L} = \frac{1}{2} \, g_{\mu\nu}^{\rm eff} (x^{\lambda}) \, \frac{dx^{\mu}}{d\tau_{\rm eff}} \, \frac{dx^{\nu}}{d\tau_{\rm eff}} - \frac{1}{2} \, \mu^2 \, .
\end{equation}
Indeed, the critical points of the action (\ref{eq5.1}), (\ref{eq5.2}) are geodesics of $g_{\mu\nu}^{\rm eff}$. The value of the Lagrangian ${\mathcal L}$ is a constant of motion which can be (after variation) constrained to have the value $-\mu^2$, corresponding to
\begin{equation}
\label{eq5.3}
\dot x^2 \equiv g_{\mu\nu}^{\rm eff} \, \frac{dx^{\mu}}{d\tau_{\rm eff}} \, \frac{dx^{\nu}}{d\tau_{\rm eff}} = - \mu^2 \, .
\end{equation}

The constraint (\ref{eq5.3}) corresponds to using as (affine) parameter $\tau_{\rm eff}$ along the geodesic the effective proper time divided by the (reduced) mass $\mu$,
\begin{equation}
\label{eq5.4}
d\tau_{\rm eff} = \frac{ds_{\rm eff}}{\mu} \, ,
\end{equation}
so that the critical value of the action (\ref{eq5.1}) coincides with the usual ``square-root'' one $- \int \mu \, ds_{\rm eff}$. [For simplicity, we by-passed here the justification of the quadratic action (\ref{eq5.2}) based on the (Polyakov-like) use of an independent ``einbein'' degree of freedom along the worldline.]

\smallskip

Inserting  the effective EOB metric (\ref{eq3.2}), one finds that the explicit form of the Lagrangian (\ref{eq5.2}) reads
\begin{equation}
\label{eq5.5}
{\mathcal L} = - \frac{1}{2} \, A(r) \left( \frac{dt}{d\tau_{\rm eff}} \right)^2 + \frac{1}{2} \, \bar B (r) \left( \frac{dr}{d\tau_{\rm eff}} \right)^2 + \frac{1}{2} \, r^2 \left( \frac{d\varphi}{d\tau_{\rm eff}} \right)^2 - \frac{1}{2} \, \mu^2 \, .
\end{equation}
The Lagrangian (\ref{eq5.5}) admits the following conserved quantities
\begin{eqnarray}
\label{eq5.6}
{\mathcal E}_{\rm eff} &= &- \frac{\partial \, {\mathcal L}}{\partial (dt / d\tau_{\rm eff})} = A(r) \, \frac{dt}{d\tau_{\rm eff}} \, , \\
\label{eq5.7}
J &= &\frac{\partial \, {\mathcal L}}{\partial (d\varphi / d\tau_{\rm eff})} = r^2 \, \frac{d\varphi}{d\tau_{\rm eff}} \, , \\
\label{eq5.8}
{\mathcal L} &= &-\mu^2 \, .
\end{eqnarray}
Inserting (\ref{eq5.6}) and  (\ref{eq5.7}) into (\ref{eq5.8}) then leads to the following equation ruling the radial motion
\begin{equation}
\label{eq5.9}
{\mathcal E}_{\rm eff}^2 = A(r) \, \bar B (r) \left( \frac{dr}{d\tau_{\rm eff}} \right)^2 + A(r) \left( \mu^2 + \frac{J^2}{r^2} \right) \, .
\end{equation}

It is easily checked that the radial dynamics defined by (\ref{eq5.9}) is equivalent to the small-eccentricity limit of the (squared) radial effective Hamiltonian (\ref{eq3.6}). [The conserved effective angular momentum $J$ having the same meaning in (\ref{eq3.6}) and (\ref{eq5.9}), and the conserved effective energy ${\mathcal E}_{\rm eff}$ being numerically equal to the effective Hamiltonian $H_{\rm eff}$.] Passing to the dimensionless rescaled variables (\ref{eq3.8}), together with a corresponding (dimensionless) rescaled effective proper time $\hat\tau$, such that
\begin{equation}
\label{eq5.10}
\tau_{\rm eff} = \frac{s_{\rm eff}}{\mu} = \frac{GM}{\mu} \, \hat\tau \, ,
\end{equation}
we find
\begin{equation}
\label{eq5.11}
\hat{\mathcal E}_{\rm eff}^2 = A(\hat r) \, \bar B (\hat r) \left( \frac{d\hat r}{d\hat\tau} \right)^2 + A(\hat r) \left( 1 + \frac{j^2}{\hat r^2} \right) \, .
\end{equation}
The rewriting of (\ref{eq5.11}) in terms of the useful EOB radial variable $u \equiv 1/\hat r \equiv GM/r$, yields the form
\begin{equation}
\label{eq5.12}
\hat{\mathcal E}_{\rm eff}^2 = \frac{A(u) \, \bar B(u)}{u^4} \left( \frac{du}{d\hat\tau} \right)^2 + A(u) (1+j^2 \, u^2) \, .
\end{equation}
The result (\ref{eq5.12}) exhibits how the circular-orbit effective potential $W_j (u)$, Eq. (\ref{eq4.2}), which depends only on the EOB potential $A(u)$, gets modified by a radial-energy term which involves the {\it product} of the two functions $A(u)$ and $\bar B (u)$, i.e. the combination $D(u)$ introduced in Eq.~(\ref{eq3.12}).

\smallskip

Small-eccentricity orbits are solutions $u = u(\hat\tau)$ of the constraint (\ref{eq5.12}) of the form
\begin{equation}
\label{eq5.13}
u(\hat\tau) = u_0 + \varepsilon \, u_1 (\hat \tau) + {\mathcal O} (\varepsilon^2) \, ,
\end{equation}
where $u_0$ is independent of $\hat\tau$, and where $\varepsilon$ is a small parameter measuring the eccentricity of the orbit. Inserting (\ref{eq5.13}) into (\ref{eq5.12}), one finds, at order $\varepsilon^0$, that $u_0$ must be an {\it extremum} of $W_j (u) \equiv A(u) (1+j^2 \, u^2) \equiv A(u) + j^2 \, B(u)$, i.e. that $u_0$ is related to $j$ by
\begin{equation}
\label{eq5.14}
j^2 = - \frac{A' (u_0)}{B' (u_0)} \, .
\end{equation}
We thereby recover that $u(\hat\tau) = u_0$ corresponds to a circular orbit (see Eqs.~(\ref{eq4.4}), (\ref{eq4.5}) above).

\smallskip

Then at order $\varepsilon^2$ (and modulo ${\mathcal O} (\varepsilon^3)$ corrections), one finds that
\begin{equation}
\label{eq5.15}
\frac{A(u_0) \, \bar B (u_0)}{u_0^4} \left( \frac{du_1}{d\hat\tau} \right)^2 + \frac{1}{2} \, W''_j (u_0) \, u_1^2 = {\rm const.} + {\mathcal O} (\varepsilon) \, ,
\end{equation}
where $W''_j \equiv \partial^2 \, W_j (u) / \partial \, u^2$ (keeping $j$ fixed). The constraint (\ref{eq5.15}) shows that $u_1 (\hat\tau)$ undergoes harmonic motion, i.e. $u_1 (\hat\tau) = \alpha \cos (\hat\omega_r \, \hat\tau + \varphi_0)$. Here $\hat\omega_r$ denotes the frequency (measured in $\hat\tau$ time) of the small {\it radial oscillations} described by $u = GM/r = u_0 + \varepsilon \, u_1 (\hat\tau) + {\mathcal O} (\varepsilon^2)$. One sees on (\ref{eq5.15}) that $\hat\omega_r^2$ is given by
\begin{equation}
\label{eq5.16}
\hat\omega_r^2 (u_0) = \frac{1}{2} \, \frac{u_0^4}{D(u_0)} \, W''_j (u_0) \, ,
\end{equation}
where we used the notation (\ref{eq3.12}). More precisely, as long as $u_0$ is a {\it minimum} of the effective potential $W_j (u)$, i.e. as long as $W''_j (u_0) > 0$, one finds that $\hat\omega_r^2$ is positive, which indeed corresponds to (stable) harmonic radial oscillations $u_1 (\hat\tau) \propto \cos (\hat\omega_r \, \hat\tau + \varphi_0)$. However, when $u_0$ is a {\it maximum} of $W_j (u)$ ($W''_j (u_0) < 0$ so that $\hat\omega_r^2 < 0$) the perturbations $\delta u = u_1 (\hat\tau)$ of the circular orbit $u=u_0$ are {\it unstable}, and grow exponentially: $u_1 (\hat\tau) = \alpha \exp (\vert \hat\omega_r \vert \, \hat\tau) + \beta \exp (-\vert \hat\omega_r \vert \, \hat\tau)$, where $\vert \hat\omega_r \vert = \sqrt{-\hat\omega_r^2}$. The dividing line is the LSO where $W''_j (u_0) = 0$, i.e. $\hat\omega_r^2 (u_0) = 0$, in agreement with the criterion (\ref{eq4.13}) used above.

\smallskip

We cannot directly relate the (EOB) $\hat\tau$-time (squared) radial frequency $\hat\omega_r^2$ to the radial frequency $\omega_r^2$ studied in the recent GSF study \cite{Barack:2009ey} because the latter quantity refers to a different time variable $\tau$, namely the proper time of the Schwarzschild background (of mass $m_2$) used in GSF work. In order to transform the result (\ref{eq5.16}) into a physical, gauge-invariant result, the simplest is to consider the {\it dimensionless} ratio of two frequencies: the radial frequency $\hat\omega_r$, and the corresponding $\hat\tau$-time azimuthal (or circular) frequency, say $\hat{\hat\Omega} = d\varphi / d\hat\tau$. [We use a double hat on $\Omega$ to distinguish it from the dimensionless {\it coordinate-time} azimuthal frequency $\hat\Omega^{\rm circ}$ defined in Eq.~(\ref{eq4.9}).] Using the angular momentum conservation law (\ref{eq5.7}), we see that, along the circular orbit $u=u_0$, the $\hat\tau$-time azimuthal frequency is given by
\begin{equation}
\label{eq5.17}
\hat{\hat\Omega} (u_0) = \frac{j}{\hat r_0^2} = j \, u_0^2 \, .
\end{equation}

Finally, dividing (\ref{eq5.16}) by the square of (\ref{eq5.17}) leads to
\begin{equation}
\label{eq5.18}
\frac{\hat\omega_r^2 (u_0)}{\hat{\hat\Omega}^2 (u_0)} = \frac{1}{2} \, \frac{1}{D(u_0)} \left( \frac{A'' (u_0)}{j^2 (u_0)} + B'' (u_0) \right) \, ,
\end{equation}
where $j^2 (u_0)$ is given by Eq.~(\ref{eq5.14}). Inserting the latter equation yields the explicit form
\begin{equation}
\label{eq5.19}
\left( \frac{\hat\omega_r}{\hat{\hat\Omega}} \right)^2 (u_0) = \frac{\bar D (u_0)}{2} \, \frac{\Delta (u_0)}{A'(u_0)} \, ,
\end{equation}
where $\Delta (u)$ is the determinant (\ref{eq4.13}) introduced above, and where $\bar D (u)$ denotes, as in (\ref{eq3.13}), the {\it inverse} of $D(u)$.

\smallskip

The quantity on the L.H.S. of (\ref{eq5.19}) is gauge-invariant, and independent of the time parametrization. In particular, it is also equal to the square of the ratio of the coordinate-time radial frequency, say $\omega_r$, to the coordinate-time azimuthal frequency $\Omega = d\varphi / dt$. The R.H.S. is expressed in terms of the EOB (inverse) radial coordinate $u_0 = GM/r_0$ (along circular orbits). However, Eq.~(\ref{eq4.10}) has related $u_0$ to the dimensionless, gauge-invariant frequency parameter $\hat\Omega_{\rm circ} (u_0) \equiv GM \, \Omega^{\rm circ} (u_0) \equiv GM \, d\varphi / dt$. Therefore, the combination of (\ref{eq4.10}) and (\ref{eq5.19}) relates the two gauge-invariant observables $(\omega_r / \Omega)^2$ and $GM \, \Omega$.

\smallskip

The EOB predictions (\ref{eq4.10}) and (\ref{eq5.19}) are valid for an arbitrary mass ratio $\nu$. Let us now focus on the limit $\nu \ll 1$, in which GSF studies can, in principle\footnote{Though there are some technical complications for dealing with unstable orbits (e.g. with extracting the conservative part of the dynamics), GSF studies can, in principle, compute the L.H.S. of (\ref{eq5.19}) for $0 < u_0 < 1/3$, i.e. for $r_0 > 3 \, GM$.}, compute the L.H.S. of (\ref{eq5.19}) as a function of $\hat\Omega = GM \, \Omega$.

\smallskip

Inserting in (\ref{eq5.19}) the GSF expansions (\ref{eq3.16a}), (\ref{eq3.16b}) one finds (suppressing for simplicity the hats, as $\hat\omega_r / \hat{\hat\Omega} \equiv \omega_r / \Omega$: the ratio of coordinate-time frequencies)
\begin{eqnarray}
\label{eq5.20}
\left( \frac{\omega_r}{\Omega} \right)^2 &= &1-6u + \nu \left[(1-6u) \, \bar d (u) + a(u) + 2u \, a'(u) + \frac{1}{2} \, u(1-2u) \, a'' (u) \right] \nonumber \\
&+ &{\mathcal O} (\nu^2) \, ,
\end{eqnarray}
where we henceforth also suppress the index $0$ on $u_0$. By further inserting in (\ref{eq5.20}) the GSF expansion of $u$ in terms of $x$, Eq.~(\ref{eq4.22}), we get
\begin{equation}
\label{eq5.21}
\left( \frac{\omega_r}{\Omega} \right)^2 = 1-6x+\nu \, \rho(x)+{\mathcal O}(\nu^2) \, ,
\end{equation}
where the function $\rho (x)$ is made of three pieces:
\begin{equation}
\label{eq5.22}
\rho (x) \equiv \rho_E (x) + \rho_a (x) + \rho_d (x) \, .
\end{equation}
The first piece, $\rho_E (x)$, comes from the Energy map (\ref{eq3.1}) relating the EOB effective energy ${\mathcal E}_{\rm eff} = H_{\rm eff}$ to the EOB real energy ${\mathcal E}_{\rm real} = H_{\rm EOB}$. It is given by
\begin{equation}
\label{eq5.23}
\rho_E (x) = 4x \left( 1-\frac{1-2x}{\sqrt{1-3x}} \right) \, .
\end{equation}
The second piece, $\rho_a (x)$, is related to the function $a(u)$. It reads
\begin{equation}
\label{eq5.24}
\rho_a (x) = a(x) + x \, a' (x) + \frac{1}{2} \, x (1-2x) \, a'' (x) \, .
\end{equation}
Finally, the third piece, $\rho_d (x)$, is related to the function $\bar d (u)$, and reads
\begin{equation}
\label{eq5.25}
\rho_d (x) = (1-6x) \, \bar d (x) \, .
\end{equation}

Eqs.~(\ref{eq5.21})--(\ref{eq5.25}) represent one of the main new results of the present work. They show that GSF studies can give us access to a function of $x \equiv G(m_1 + m_2) \, \Omega$ which, when interpreted within the EOB formalism, is an $x$-dependent combination of $a(x)$, $a'(x)$, $a''(x)$ and $\bar d (x)$, where $a(x)$ and $\bar d (x)$ are the $\nu$-derivatives of the two basic functions $A(u;\nu)$ and $\bar D (u;\nu)$ parametrizing the EOB effective metric. In principle, GSF calculations can numerically compute the function $\rho (x)$ on the full interval $0 < x < 1/3$ corresponding to $3 \, GM < r_0 < +\infty$.

\smallskip

Note that the contribution $\rho_d (x)$ contains a factor $1-6x$ which vanishes at the (unperturbed) LSO. This implies that the value of the ($\nu$-perturbed) LSO frequency obtained by requiring the vanishing of $(\omega_r / \Omega)^2$ is equivalent to that obtained by discarding the $\rho_d (x)$ contribution to $\rho (x)$. More precisely, the solution of $(\omega_r / \Omega)^2 = 0$ is
\begin{eqnarray}
\label{eq5.26}
x_{\rm LSO} &= &\frac{1}{6} \left[ 1+\nu \, \rho ( \mbox{$\frac{1}{6}$} ) + {\mathcal O} (\nu^2) \right] \nonumber \\
&= &\frac{1}{6} \left[ 1+\nu \left( \rho_E ( \mbox{$\frac{1}{6}$} ) + \rho_a ( \mbox{$\frac{1}{6}$} ) \right) + {\mathcal O} (\nu^2) \right] \, ,
\end{eqnarray}
which is easily seen to coincide with our previous result (\ref{eq4.26})--(\ref{eq4.28}).

\smallskip

On the other hand, though the {\it exact} ($\nu$-linear) result (\ref{eq5.21}) is equivalent to the {\it exact} ($\nu$-linear) result (\ref{eq4.26})--(\ref{eq4.28}) for determining the {\it exact} ($\nu$-linear) $x_{\rm LSO}$, the {\it PN expansion} of (\ref{eq5.21}) leads to a different estimate of the {\it PN expansion} of $x_{\rm LSO}$ than the {\it PN expansion} of (\ref{eq4.26}). Indeed, let us consider the PN expansion (i.e. the expansion in powers of $x$) of the exact function $\rho (x)$:
\begin{equation}
\label{eq5.27}
\rho^{\rm PN} (x) = \rho_E^{\rm PN} (x) + \rho_a^{\rm PN} (x) + \rho_d^{\rm PN} (x) \, ,
\end{equation}
where
\begin{equation}
\label{eq5.28}
\rho_E^{\rm PN} (x) = 2 \, x^2 - \frac{3}{2} \, x^3 - \frac{27}{4} \, x^4 - \frac{675}{32} \, x^5 - \frac{3969}{64} \, x^6 + {\mathcal O} (x^7) \, ,
\end{equation}
\begin{eqnarray}
\label{eq5.29}
\rho_a^{\rm PN} (x) &= &3 \, a_3 \, x^2 + (6 \, a_4 - 2 \, a_3) \, x^3 + (10 \, a_5 - 7 \, a_4) \, x^4  \nonumber \\
&+ &(15 \, a_6 - 14 \, a_5) \, x^5 +(21 \, a_7 - 23 \, a_6) \, x^6 + {\mathcal O} (x^7) \, ,
\end{eqnarray}
\begin{eqnarray}
\label{eq5.30}
\rho_d^{\rm PN} (x) &= &\bar d_2 \, x^2 + (\bar d_3 - 6 \, \bar d_2) \, x^3 + (\bar d_4 - 6 \, \bar d_3) \, x^4  \nonumber \\
&+ &(\bar d_5 - 6 \, \bar d_4) \, x^5 + (\bar d_6 - 6 \, \bar d_5) \, x^6 + {\mathcal O} (x^7) \, .
\end{eqnarray}
Here the coefficients $a_n$ ($n \geq 3$) are the expansion coefficients of the PN expansion (\ref{eq4.31}) of the function $a(u)$. Similarly the coefficients $\bar d_n$ ($n \geq 2$) are the expansion coefficients of the PN expansion of the function $\bar d (u)$, i.e.
\begin{equation}
\label{eq5.31}
\bar d (u) = \sum_{n \geq 2} \bar d_n \, u^n \, .
\end{equation}
As above, we are here assuming, for simplicity, that the expansion coefficients $a_n$ and $\bar d_n$ are pure numbers (i.e. we neglect any possible logarithmic running starting at the 4PN level: $a_5 , \bar d_4$). Combining Eqs.~(\ref{eq5.28})--(\ref{eq5.30}) one finds for the PN expansion coefficients of $\rho (x)$, i.e.
\begin{equation}
\label{eq5.32}
\rho^{\rm PN} (x) = \rho_2 \, x^2 + \rho_3 \, x^3 + \rho_4 \, x^4 + \rho_5 \, x^5 + \rho_6 \, x^6 + \ldots
\end{equation}
\begin{eqnarray}
\label{eq5.33a}
\rho_2 &= &2+3 \, a_3 + \bar d_2 \, , \\
\label{eq5.33b}
\rho_3 &= &-\frac{3}{2} - 2 \, a_3 + 6 \, a_4 - 6 \, \bar d_2 + \bar d_3 \, , \\
\label{eq5.33c}
\rho_4 &= &-\frac{27}{4} - 7 \, a_4 + 10 \, a_5 - 6 \, \bar d_3 + \bar d_4 \, , \\
\label{eq5.33d}
\rho_5 &= &-\frac{675}{32} - 14 \, a_5 + 15 \, a_6 - 6 \, \bar d_4 + \bar d_5 \, , \\
\label{eq5.33e}
\rho_6 &= &-\frac{3969}{64} - 23 \, a_6 + 21 \, a_7 - 6 \, \bar d_5 + \bar d_6 \, .
\end{eqnarray}

The 3PN-level calculation of the dynamics of two-body systems \cite{Damour:2001bu} has given us access to the values of $a_3$ and $a_4$ (see (\ref{eq4.34})), as well as of $\bar d_2$ and $\bar d_3$ \cite{Damour:2000we}, namely
\begin{equation}
\label{eq5.34}
\bar d_2 = 6 \, , \qquad \bar d_3 = 52 \, .
\end{equation}

Note that \cite{Damour:2000we} has also given us access to the beginning of the PN expansions of the $\nu$-quadratic (2GSF) (and even $\nu$-cubic, 3GSF, for $A(u;\nu)$) terms in the GSF expansions (\ref{eq3.16a}), (\ref{eq3.16b}) of $A(u;\nu)$ and $\bar D (u;\nu)$. Namely $a_2^{3{\rm PN}} (u) = a_3^{3{\rm PN}} (u) = 0$, and $\bar d_2^{3{\rm PN}} (u) = -6 \, u^3$ (see Eqs.~(\ref{eq3.15a}), (\ref{eq3.15b})). [We recall that the 3PN expansion of $A(u)$ goes through the order $u^4$, while that of $\bar D(u)$ stops at order $u^3$.]

\smallskip

Using the knowledge of $a_3$, $a_4$ (Eqs. (\ref{eq4.34})) and $\bar d_2 , \bar d_3$ (Eqs.~(\ref{eq5.34})) we can compute the first two terms in the PN expansion (\ref{eq5.32}) of $\rho (x)$, namely
\begin{eqnarray}
\label{eq5.35}
\rho_2 &= &14 \, , \nonumber \\
\rho_3 &= &\frac{397}{2} - \frac{123}{16} \, \pi^2 = 122.627416 \, .
\end{eqnarray}
We have checked that the corresponding 1GSF 3PN-accurate expansion of the ratio $(\omega_r / \Omega)^2$, (\ref{eq5.21}), agrees with the (full GSF, 3PN-accurate) results derived by Damour, Jaranowski and Sch\"afer \cite{Damour:1999cr}. [Note that one must use the values $\omega_{\rm kinetic} = 41/24$ \cite{Damour:2000kk,Blanchet:2000nv} and $\omega_s = 0$ \cite{Damour:2001bu} in the results of \cite{Damour:1999cr}.] Indeed these authors determined the PN expansions of both
$$
\omega_r \equiv \omega_{\rm radial} \equiv 2\pi / P
$$
and 
$$
\Omega \equiv \omega_{\rm circ} \equiv \omega_{\rm radial} + \omega_{\rm periastron} = 2\pi \, \frac{1+k}{P} \, .
$$
The ratio $(\omega_r / \Omega)^2$ corresponds, in their notation, to $(1+k)^{-2}$. For circular orbits, Ref.~\cite{Damour:1999cr} computed the expansion in powers of $1/j^2$ of $k$ (and $(1+k)^{-4}$), see their Eqs~(5.25)--(5.28). Inserting in these expansions their result (5.8) for the $x$ expansion of $j_{\rm circ}$ one finds the $\nu$-exact, 3PN-accurate expansion of $(\omega_r / \Omega)^2$ in powers of $x$. Its $\nu$-linear piece is found to agree with our result (\ref{eq5.35}) above.

\smallskip

If one were to use the 3PN expansion of $\rho (x)$ to estimate the LSO frequency, i.e. from the {\it first} equation (\ref{eq5.26}) (keeping the $d$-contribution), one would find
\begin{equation}
\label{eq5.36}
x_{\rm LSO} [\rho^{3{\rm PN}}] = \frac{1}{6} \, [1+c_x [\rho^{3{\rm PN}}] \, \nu + {\mathcal O} (\nu^2)]
\end{equation}
with
\begin{equation}
\label{eq5.37}
c_x [\rho^{3{\rm PN}}] = \frac{\rho_2}{6^2} + \frac{\rho_3}{6^3} = 0.956608 \, .
\end{equation}
If we compare this estimate to the previous 3PN-based estimate (\ref{eq4.38}) (based on the PN expansion of the exact, $a(u)$-dependent, result (\ref{eq4.26})), as well as to the recent GSF estimate (\ref{eq2.20}), we see that (\ref{eq5.37}) is now {\it larger} (by $14.67\%$) than the GSF estimate, and {\it larger} (by $58.26\%$) than the previous $a^{3{\rm PN}}$ estimate (\ref{eq4.38}). This shows the unreliability of using (non resummed) PN expansions for estimating physical quantities in the strong-field regime (here the LSO, which is a non-perturbative phenomenon). Although the $\rho^{3{\rm PN}}$ estimate (\ref{eq5.37}) happens to be closer to the GSF result (\ref{eq2.20}) than the $a^{3{\rm PN}}$ one (\ref{eq4.38}), we think that this is purely accidental. Indeed, the results of Section~4 have shown that $c_x^{\rm EOB}$ only depends on the $a(u)$ function, and that the present $27.5\%$ disagreement (on the low side) between $c_x [a^{3{\rm PN}}]$ and $c_x^{\rm GSF}$ is most probably due to neglecting the higher PN contributions to the $a$ function, see Eq.~(\ref{eq4.40}). By contrast, when comparing the various terms contribution to $c_x [\rho^{3{\rm PN}}]$, (\ref{eq5.37}), one finds that it contains a rather large contribution connected with the large value of $\bar d_3$ which ``pollutes'' $\rho_3$, see Eq.~(\ref{eq5.33b}). Indeed, note that the individual $d$-contributions to $c_x[\rho^{3{\rm PN}}]$ are
\begin{eqnarray}
\label{eq5.38}
\rho_{d2} \, \frac{1}{6^2} &= &\frac{\bar d_2}{6^2} \, , \nonumber \\
\rho_{d3} \, \frac{1}{6^3} &= &-\frac{\bar d_2}{6^2} + \frac{\bar d_3}{6^3} \, ,
\end{eqnarray}
so that their sum yields
\begin{equation}
\label{eq5.39}
\rho_{d2} \, \frac{1}{6^2} + \rho_{d3} \, \frac{1}{6^3} = \frac{\bar d_3}{6^3} = \frac{52}{6^3} = 0.24074074 \, .
\end{equation}

The crucial point is that while $\bar d_2$ cancells (as it should) between the 2PN and 3PN contributions to $c_x [\rho^{3{\rm PN}}]$, $\bar d_3 = +52$ does not cancell, and thereby artificially overestimates $c_x$. One sees on Eq.~(\ref{eq5.33c}) that the 4PN contribution to $c_x [\rho^{4{\rm PN}}]$ will introduce another contribution proportional to $\bar d_3$ which precisely cancells the contribution (\ref{eq5.39}). Similarly one sees on Eqs.~(\ref{eq5.33a})--(\ref{eq5.33e}) that the contribution $\propto \bar d_4$ that appears at the 4PN level, Eq.~(\ref{eq5.33c}), is cancelled by a contribution $\propto -6 \, \bar d_4$ entering the 5PN term (\ref{eq5.33d}). In other words, while the exact expression (\ref{eq5.25}) of $\rho_d (x)$ shows that $\bar d(x)$ does not influence the LSO frequency, the PN expansion (\ref{eq5.30}) of $\rho_d (x)$ artificially introduces a contribution of $\bar d(x)$ at any finite PN order.

\smallskip

A more general way of seeing the unreliability of using (non resummed) PN expansions to determine the LSO consists in considering, given any function of $x$ (which does not vanish at the LSO), say $f(x;\nu)$, the quantity
\begin{equation}
\label{eq5.40}
F(x;\nu) \equiv f(x;\nu) \left( \frac{\omega_r}{\Omega} \right)^2 \, .
\end{equation}
The {\it exact} solution of the constraint $F(x) = 0$ is always the exact LSO, independently of the multiplicative factor $f(x)$. However, if one considers the PN expansion of $F(x)$ and solves some PN-truncated equation, say $F^{n{\rm PN}} (x) = 0$, its solution will depend on the coefficients entering the PN expansion of $f(x)$.

\smallskip

Going back to the issue of the knowledge that can be extracted from GSF calculations of the function $(\omega_r / \Omega)^2 (x)$, i.e. the knowledge of the function $\rho (x)$, see Eq.~(\ref{eq5.21}), we note that it would be very interesting to study this function not only near the LSO $\left( x \sim \frac{1}{6} \right)$, but also for small values of $x$ (i.e. large radii). In this limit, one should be able to: (i) check the validity of the first two terms, $\rho_2 \, x^2 + \rho_3 \ x^3$ in the PN expansion of $\rho (x)$, and (ii) reliably extract some of the higher-order terms in the PN expansion (\ref{eq5.32}) (and explore their logarithmic running). Note in this respect that, in view of Eqs.~(\ref{eq5.33a})--(\ref{eq5.33e}), measuring higher and higher coefficients $\rho_n$ will give us access to some combinations of the higher PN coefficients of $a(x)$ and $\bar d (x)$. More precisely, we will have the structure
\begin{eqnarray}
\label{eq5.41}
\rho_4 &\to &10 \, a_5 + \bar d_4 \, , \nonumber \\
\rho_5 &\to &-14 \, a_5 + 15 \, a_6 - 6 \, \bar d_4 + \bar d_5 \, , \nonumber \\
\rho_6 &\to &-23 \, a_6 + 21 \, a_7 - 6 \, \bar d_5 + \bar d_6 \, ,
\end{eqnarray}
where the R.H.S.'s indicate the knowledge acquired from knowing the L.H.S.'s. Though it is somewhat frustrating to see the (less interesting) higher-order $\bar d_n$ coefficients entering (and ``polluting'') the R.H.S.'s we might still be able to reach some plausible results by combining (\ref{eq5.41}) with our previous result (\ref{eq4.40}) which only contained the (more interesting) $a_n$ coefficients. In particular, as it is easily checked that the main contribution to the 3PN coefficient $\rho_3$, Eq.~(\ref{eq5.35}), comes from $a_4$, it is plausible (especially in view of its rather large numerical coefficient) that the contribution $10 \, a_5$ dominates in the quantity $10 \, a_5 + \bar d_4$ which can be deduced from a measurement of $\rho_4$. Then a GSF measurement of $\rho_4$ would give us an approximate knowledge of $a_5$. Inserting this knowledge in the approximate equation (\ref{eq4.42}) above, we would then be able to approximately determine $a_6$ too. It will then be interesting to see whether the values of $a_5$ and $a_6$ so determined are close to our EOB-NR-based estimate (\ref{eq4.45}). If this is the case one might continue the approximation and try to deduce, e.g., an approximate estimate of $a_7$. In addition, GSF data on the small $x$ behaviour of $\rho (x)$ will allow one to explore the presence of logarithmic terms starting at the 4PN level (see Section~7 below where one similarly discusses how GSF data on Detweiler's redshift function $u^t (x)$ can explore the presence of logarithmic terms in the latter function).

\section{Zero-binding zoom-whirl motion and other \break ways of extracting EOB information from \break GSF computations}
\setcounter{equation}{0}

There are several other ways in which GSF computations might be used to extract information about the functions $A(u)$, $\bar D (u)$, $Q(u,p_r)$ entering the EOB formalism. Let us here sketch a few possibilities.

\smallskip

First, one might go beyond the small-eccentricity limit and compare 1GSF computations of eccentric orbits to EOB predictions. However, to do so one should include the effect of the ${\mathcal O} (p_r^4)$ contribution $Q(u,p_r)$. At present one only knows the 3PN-level expression (\ref{eq3.11}) of $Q(u,p_r) \sim u^2 \, p_r^4$. At higher orders it will probably involve both higher powers of $u$ and higher powers of $p_r$, say
\begin{equation}
\label{eq6.1}
Q(u,p_r) = \nu [q(u) \, p_r^4 + \bar q (u) \, p_r^6] + {\mathcal O} (\nu^2) + {\mathcal O} (p_r^8) \, ,
\end{equation}
with $q(u) = q_2 \, u^2 + q_3 \, u^3 + \ldots$. A detailed EOB/GSF comparison for smallish-eccentricity orbits might allow one to probe the function $q(u)$, together with $a(u)$ and $\bar d (u)$.

\smallskip

Second, a more ambitious project might be to try to extract the total conserved (EOB) energy and angular momentum, ${\mathcal E}$ and $J$ from GSF computations. In principle, one GSF way of determining ${\mathcal E}$ and ${\mathcal J}$ to the required accuracy would be to go to the {\it second} GSF approximation (2GSF), i.e. to compute the metric perturbation through order $\nu^2$, and then to compute, from the metric, the ADM surface integrals giving the {\it total} energy and angular momentum of the system. This will give ${\mathcal E} - M = e_1 \, \nu + e_2 \, \nu^2 + {\mathcal O} (\nu^3)$ and ${\mathcal J} = j_1 \, \nu + j_2 \, \nu^2 + {\mathcal O} (\nu^3)$, which is the precision needed to explore the effects linked to $\nu \, a(u)$, $\nu \, \bar d (u), \ldots$ However, it is not clear when the GSF community will be in position to compute the 2GSF approximation. Let us therefore sketch what can be done now with the 1GSF approximation. The knowledge of the 1GSF conservative self-force allows one, in principle, to determine some {\it conserved} energy-like and angular-momentum-like quantities along perturbed orbits. The clearest way to do so is to consider {\it unbound} orbits, that come from some ``in state'' with infinite separation between $m_1$ and $m_2$. The ``in state'' then determines, in principle, the numerical value of ${\mathcal E}$ and $J$. One must, however, be careful to include the effect of the recoil of the large mass $m_2$. In an Hamiltonian formalism (such as the ADM or the EOB one), the total conserved angular momentum is given by
\begin{equation}
\label{eq6.2}
{\bm J} = {\bm x}_1 \times {\bm p}_1 + {\bm x}_2 \times {\bm p}_2 \, ,
\end{equation}
while the total conserved energy has the form
\begin{equation}
\label{eq6.3}
{\mathcal E} = H = E_1^0 + E_2^0 + H_{\rm int} \, ,
\end{equation}
where $E_a^0 \equiv (m_a^2 + {\bm p}_a^2)^{1/2}$, $a = 1,2$, denotes the {\it free} kinetic mass-energy, and where the {\it interaction} term $H_{\rm int}$ tends to zero in the infinite separation limit $r \to \infty$. In addition, the conditions to be in the center of mass frame have the form 
\begin{equation}
\label{eq6.4}
{\bm p}_1 + {\bm p}_2 = 0 \, ,
\end{equation}
\begin{equation}
\label{eq6.5}
E_1^0 \, {\bm x}_1 + E_2^0 \, {\bm x}_2 + \mbox{interaction terms} = 0 \, ,
\end{equation}
where the interaction terms are proportional to the gravitational constant, and are smaller by a factor $r$ than the leading terms in (\ref{eq6.5}) (see \cite{Damour:2000kk}). Usually, GSF calculations only follow the dynamics of the small mass $m_1$. It would then record (for unbound orbits) only the $m_1$ contribution to ${\mathcal E}$ and ${\bm {\mathcal J}}$ coming from the incoming state (with infinite separation), say
\begin{equation}
\label{eq6.6}
{\bm J}_1^{\infty} = [ {\bm x}_1 \times {\bm p}_1]_{\infty} \, ,
\end{equation}
\begin{equation}
\label{eq6.7}
E_1^{\infty} = [E_1^0]_{\infty} = \left[ \sqrt{m_1^2 + {\bm p}_1^2} \right]_{\infty} \, .
\end{equation}

In view of (\ref{eq6.2})--(\ref{eq6.5}) one must then ``renormalize'' both the angular momentum and the energy
\begin{equation}
\label{eq6.8}
{\bm J} = \left( 1+\frac{E_1^{\infty}}{E_2^{\infty}} \right) {\bm J}_1^{\infty} \, ,
\end{equation}
\begin{equation}
\label{eq6.9}
{\mathcal E} = E_1^{\infty} + \left[ \sqrt{m_2^2 + {\bm p}_1^2} \right]_{\infty} \, .
\end{equation}

Having so computed ${\mathcal E}$ and $J$ one might then, for instance, compare the EOB prediction for the scattering angle $\theta ({\mathcal E} , J)$ (which follows from the EOB Hamiltonian) with GSF computations of $\theta$ for a sample of values of ${\mathcal E}$ and ${\mathcal J}$. We see that, in principle, we have here access to one function of {\it two} real variables, which is ample information for determining the functions entering the EOB formalism.

\smallskip

We have been assuming here that, with some effort (notably concerning the separation of the {\it conservative} part of the self-force from its radiative part), one could deal with unbound orbits. In case it is computationally much easier to deal with bound orbits, one might still be able to extract ${\mathcal E}$ and $J$ from GSF computations. First, we remark that in view of the parity properties of the {\it conservative} self-force under time reversal (see, e.g., \cite{Barack:2009ey}) the {\it ratios} $R_t \equiv F_t / u^r$ and $R_{\varphi} \equiv F_{\varphi} / u^r$ (where $u^r = dr / d\tau$, using the GSF proper time of, say, \cite{Barack:2009ey}) are {\it even} functions of $\tau$. [Here, we fix $\tau = 0$ at an extremum of the radial coordinate.] This even-parity property is shared by the radius as a function of $\tau$. We can then follow a strategy used in \cite{Barack:2009ey} and (numerically) consider that the ratios $R_t$ and $R_{\varphi}$ are some functions of $r$.

\smallskip

This leads to evolution equations for the ``$m_1$ contribution'' to the energy and angular momentum of the form
\begin{equation}
\label{eq6.10}
\frac{d}{d\tau} \, (m_1 \, u_t) = F_t = \frac{dr}{d\tau} \, R_t (r;E_1^0 , L_1^0) \, ,
\end{equation}
\begin{equation}
\label{eq6.11}
\frac{d}{d\tau} \, (m_1 \, u_{\varphi}) = F_{\varphi} = \frac{dr}{d\tau} \, R_{\varphi} (r;E_1^0 , L_1^0) \, .
\end{equation}
Multiplying both sides of these equations by $d\tau$ one sees that the consideration of the radial integrals $\int dr \, R_t (r)$ and $\int dr \, R_{\varphi} (r)$ allows one to ``correct'' the ``bare particle energy'' $-m_1 \, u_t$ and the ``bare particle angular momentum'' by additional functions of $r$ so as to define some 1GSF-conserved ``improved'' particle energy and angular momentum. By exploring the $r$-dependence of these improved quantities as $r$ gets large, one might be able to so deduce the quantities $J_1^{\infty}$ and $E_1^{\infty}$, Eqs.~(\ref{eq6.6}), (\ref{eq6.7}), i.e. the values of the $m_1$ contributions for infinite separation. It then remains to ``correct'' them so that, as in Eqs.~(\ref{eq6.8}), (\ref{eq6.9}) above, they also incorporate the ``recoil'' contribution (or, more precisely, the analytic continuation in ${\mathcal E}$ and ${\mathcal J}$ of the functions that express these recoil properties in the infinite-separation limit). Finally, if one succeeds in determining ${\mathcal E}$ and ${\mathcal J}$ for bound orbits, the consideration, say as in Ref.~\cite{Damour:1999cr}, of the {\it two} gauge-invariant functions of {\it two} gauge-invariant variables
\begin{equation}
\label{eq6.12}
\omega_{\rm radial} = \omega_r ({\mathcal E} , J) \, ,
\end{equation}
\begin{equation}
\label{eq6.13}
\omega_{\rm periastron} = \omega_p ({\mathcal E} , J) \, ,
\end{equation}
will give us access to ample information for determining the functions $a(u)$, $\bar d(u)$, $Q(u,p_r)$ entering the EOB formalism.

\smallskip

Let us end this Section by considering a special motion which should be easier to investigate by GSF means, and which should give us access to very useful information about the crucial EOB function $a(u)$, independently from the other ingredients of the EOB formalism. Specifically, we wish to consider here the special {\it zero-binding zoom-whirl} orbit which starts, in the infinite part, with zero kinetic energy (but a non-zero angular momentum) at infinite separation and ends up, in the infinite future, ``whirling'' indefinitely around some limiting finite separation. In the Schwarzschild case ($\nu \to 0$ limit) this motion has dimensionless angular momentum $j_* = 4$ and its limiting dimensionless whirl radius is $\hat r_{\rm whirl} = 4$. Our purpose is to study the EOB predictions for the modifications of these values when $\nu \ne 0$. From the GSF point of view, once one has a way to compute the {\it conservative} part of the self-force (for such an unbound orbit), it should be straightforward to determine which deviation from $u_{\varphi} = 4$ is needed to end up whirling undefinitely around the large mass. Having so determined both $j_*$ (taking into account the ``correction'' (\ref{eq6.8})), and $\hat r_{\rm whirl}$, one will also have access to the final orbital frequency $\hat\Omega_{\rm whirl}$ corresponding to $\hat r_{\rm whirl}$. In other words, this special zero-binding zoom-whirl motion gives us access to {\it two} dimensionless observables: $j_*$ and $\hat\Omega_{\rm whirl}$. Then, as discussed below, one can also measure a {\it third} dimensionless observable: the value of $\omega_r^2 / \Omega^2$ at the whirl radius. 

\smallskip

From the EOB point of view the special motion we are considering has a total energy ${\mathcal E} = H_{\rm EOB}$ equal to the total mass $M$. From Eq.~(\ref{eq3.1}) we see that this ``zero-binding condition'' is equivalent to requiring that the dimensionless {\it effective} energy $\hat{\mathcal E}_{\rm eff} = \hat H_{\rm eff}$ be equal to 1. In the EOB formalism a generic radial motion is described by the constraint
\begin{equation}
\label{eq6.14bis}
W_j (u) + A(u) \left[ \frac{\hat p_r^2}{\bar B(u)} + \hat Q (u,\hat p_r) \right] = \hat{\mathcal E}_{\rm eff}^2 = {\rm const} \, ,
\end{equation}
where $u \equiv 1/\hat r$, and where
\begin{equation}
\label{eq6.15}
W_j (u) \equiv A(u) \, (1+j^2 \, u^2)
\end{equation}
is the ``effective radial potential''. Among the motions described by (\ref{eq6.14bis}) we are interested in the special case where $\hat r$ starts, in the infinite past, at $\hat r = + \infty$ with zero radial velocity (and therefore zero radial momentum $\hat p_r$), and ends, in the infinite future, at some limiting radius $\hat r_{\rm whirl}$, with zero radial velocity again. It is easily seen that these two conditions amounts to saying that the ``horizontal'' level line $\hat{\mathcal E}_{\rm eff}^2 = 1$, in the plane $(\hat r , \hat{\mathcal E}_{\rm eff}^2)$ [or $(u,\hat{\mathcal E}_{\rm eff})$], must satisfy the following conditions: (i) it intersects the ``effective potential curve'' $\hat{\mathcal E}_{\rm eff}^2 = W_j (u)$, both at $\hat r = \infty$ and at $\hat r = \hat r_{\rm whirl}$; and (ii) it is {\it tangent} to the effective potential curve at $\hat r = \hat r_{\rm whirl}$. The condition (i) expresses the fact that $\hat r = \infty$ (or $\hat u = 0$) and ${\hat r} = \hat r_{\rm whirl}$ (or $\hat u = \hat u_{\rm whirl} \equiv 1/\hat r_{\rm whirl}$) are {\it turning points} of the radial motion $(\hat p_r = 0)$, while the condition (ii) expresses the fact that the formal turning point $\hat r = \hat r_{\rm whirl}$ is reached only after an infinite time. In other words, (ii) expresses the fact that $\hat r = \hat r_{\rm whirl}$ is an {\it unstable circular orbit}. Note also that, from Eq.~(\ref{eq3.1}), the condition ${\mathcal E}_{\rm eff} = 1$ does correspond to ${\mathcal E} = H_{\rm EOB} = M$ (``zero binding''). In terms of equations, (i) and (ii) imply that the special value $j_*$ corresponding to the zero-binding zoom-whirl motion must satisfy
\begin{equation}
\label{eq6.16}
W_{j_*} (u_{\rm whirl}) = A (u_{\rm whirl}) + j_*^2 \, B (u_{\rm whirl}) = 1 \, ,
\end{equation}
\begin{equation}
\label{eq6.17}
W'_{j_*} (u_{\rm whirl}) = A' (u_{\rm whirl}) + j_*^2 \, B' (u_{\rm whirl}) = 0 \, .
\end{equation}

Note the fact that the conditions (\ref{eq6.16}), (\ref{eq6.17}) for the special zero-binding zoom-whirl motion depends {\it only} on the $A(u;\nu)$ function (remembering that $B(u)$ is just a short-hand notation for $u^2 A(u)$). In particular, it does not depend on the higher-order $\hat Q (u,\hat p_r)$ term in (\ref{eq6.14bis}). Indeed, though we do not know this term beyond the 3PN approximation (\ref{eq3.11}), it is enough to know that $\hat Q (u,\hat p_r)$ exactly vanishes as ${\mathcal O} (\hat p_r^4)$ as $\hat p_r \to 0$. [Actually even a weaker vanishing $\propto \hat p_r^2$ would suffice.] This remarkable property of this special motion makes it especially worth to be explored by a GSF approach.

\smallskip

Eliminating $j_*^2$ from these two equations (which is equivalent to using the circularity condition (\ref{eq4.5})) leads to the condition
\begin{equation}
\label{eq6.18}
A \, B' - A' \, B = B' \, .
\end{equation}
Remembering that $B(u)$ is just a short-hand notation for $u^2 \, A(u)$, the condition (\ref{eq6.18}) can be written down more explicitly as
\begin{equation}
\label{eq6.19}
A^2 = A + \frac{1}{2} \, u \, A' \, .
\end{equation}
In Eqs.~(\ref{eq6.18}) and (\ref{eq6.19}) we have provisionally suppressed the mention that the argument $u$ must take the special value $u=u_{\rm whirl}$.

\smallskip

So far we have made no approximations. Let us now consider the 1GSF approximation to the exact conditions (\ref{eq6.17}), (\ref{eq6.19}), i.e. let us insert in them $A(u;\nu) = 1-2u + \nu \, a(u) + {\mathcal O} (\nu^2)$. The 1GSF expansion of (\ref{eq6.19}) reads
\begin{equation}
\label{eq6.20}
u \, (4u -1) + (1-4u) \, \nu \, a(u) = \frac{1}{2} \, \nu \, u \, a'(u) \, .
\end{equation}
In the $\nu \to 0$ limit one recovers that its solution is $u_{\rm whirl} = \frac{1}{4} + {\mathcal O} (\nu)$, as mentioned above. Inserting this zero-th order knowledge back in (\ref{eq6.20}), one finds that the 1GSF ($\nu$-linear) modification of $u_{\rm whirl}$ is predicted, by the EOB formalism, to be
\begin{equation}
\label{eq6.21}
u_{\rm whirl} \equiv \frac{1}{\hat r_{\rm whirl}} = \frac{1}{4} \left[ 1 + \frac{\nu}{2} \, a' ( \mbox{$\frac{1}{4}$} ) + {\mathcal O} (\nu^2) \right] \, .
\end{equation}
By itself, this result is not gauge-invariant as it refers to an EOB radial coordinate. Let us now convert the result (\ref{eq6.21}) into physical, gauge-invariant results.

\smallskip

On the one hand, by inserting (\ref{eq6.21}) into either (\ref{eq6.16}) or (\ref{eq6.17}) one can compute the special total angular momentum $j_*$ of the zero-binding zoom-whirl motion. One finds
\begin{equation}
\label{eq6.22}
j_* = 4 \left[ 1-2\nu \, a ( \mbox{$\frac{1}{4}$} ) + {\mathcal O} (\nu^2)\right] \, .
\end{equation}

In addition, by inserting (\ref{eq6.21}) into our previous result  (\ref{eq4.21}), we can compute the (``flat coordinate time'') orbital frequency parameter of the limiting zero-binding circular orbit
\begin{equation}
\label{eq6.23}
x_{\rm whirl} = \frac{1}{4} \left[ 1+ \frac{\nu}{3} \, a' ( \mbox{$\frac{1}{4}$} ) + {\mathcal O} (\nu^2)\right] \, .
\end{equation}

The corresponding dimensionless frequency $\hat\Omega_{\rm whirl} = M \, \Omega_{\rm whirl} = x_{\rm whirl}^{3/2}$ is then given by
\begin{equation}
\label{eq6.24}
\hat\Omega_{\rm whirl} = \frac{1}{8} \left[ 1 + \frac{\nu}{2} \, a' ( \mbox{$\frac{1}{4}$} ) + {\mathcal O} (\nu^2)\right] \, .
\end{equation}

Note the simplifying facts that $j_*$, Eq.~(\ref{eq6.22}), depends only on the value of the $a$ function at $u = 1/4$, while $x_{\rm whirl}$ and $\hat\Omega_{\rm whirl}$, Eqs.~(\ref{eq6.23}), (\ref{eq6.24}), only depend on the value there of the derivative of $a(u)$. [The ``energy terms'' involving $(1-2u)/\sqrt{1-3u} - 1$ in (\ref{eq4.21}) have also disappeared because of the zero-binding condition.] Finally, by extending the calculations of \cite{Barack:2009ey} to unstable circular orbits, one might also have access to the value of the ratio $\omega_r^2 / \Omega^2$ (with $\omega_r^2 < 0$) at our special zero-binding limiting circular orbit (\ref{eq6.21}). In other words, one might also, using Eq.~(\ref{eq5.21}), know the value of $\rho ( \mbox{$\frac{1}{4}$} )$. Using Eqs.~(\ref{eq5.23})--(\ref{eq5.25}) and our results (\ref{eq6.22}), (\ref{eq6.23}), this will give us the value of the following combination, $\frac{1}{16} \, a'' ( \mbox{$\frac{1}{4}$} ) - \frac{1}{2} \, \bar d ( \mbox{$\frac{1}{4}$} )$, involving now the second derivative of $a(u)$ and the function $\bar d (u)$. Therefore a GSF computation of the characteristics of the special zero-binding zoom-whirl motion would give use access to two interesting pieces of information concerning the strong-field behaviour of the $a(u)$ function: $a ( \mbox{$\frac{1}{4}$} )$ and $a' ( \mbox{$\frac{1}{4}$} )$. In addition, it will give us the value of a combination of $a'' ( \mbox{$\frac{1}{4}$} )$ and $\bar d ( \mbox{$\frac{1}{4}$} )$. As the argument $u_* \simeq 1/4$ is larger than the (unperturbed) LSO value $u_{\rm LSO} \simeq 1/6$, the knowledge of $a( \mbox{$\frac{1}{4}$} )$ and $a'( \mbox{$\frac{1}{4}$} )$ will provide a strong constraint on the shape of the function $a(u)$ for all the arguments $u < \frac{1}{4}$. Note also that the PN expansion of $a( \mbox{$\frac{1}{4}$} )$ reads (neglecting any logarithmic running)
\begin{equation}
\label{eq6.25}
a ( \mbox{$\frac{1}{4}$} ) = \frac{a_3}{4^3} + \frac{a_4}{4^4} + \frac{a_5}{4^5} + \frac{a_6}{4^6} + \frac{a_7}{4^7} + \ldots
\end{equation}

If the rough estimate mentionned above, $a_n \sim \pm \, 2^n$, holds this series would behave as $\underset{n}{\sum} \pm 2^{-n}$. In particular, the $n=7$ contribution might be as small as $\pm \, 1/128$, i.e. smaller than $1\%$. Then we can use the current 3PN results (\ref{eq4.34}), together with our tentative estimate (\ref{eq4.45}) of $a_5$ and $a_6$, to ``predict'' an approximate estimate of $a( \mbox{$\frac{1}{4}$} )$, namely
\begin{eqnarray}
\label{eq6.26}
a( \mbox{$\frac{1}{4}$} ) &\simeq &0.03125 + 0.07300 - 0.02178 + 0.06152 \nonumber \\
&\simeq &0.144 \, .
\end{eqnarray}
The first line of (\ref{eq6.26}) indicates the successive contributions of $a_3$, $a_4$, $a_5$ and $a_6$. Because of the large value of $a_6^{\cap}$, (\ref{eq4.45}), which is somewhat larger than the naive expectation $\pm \, 2^6 = \pm \, 64$ (indeed $(a_6^{\cap})^{1/6} \simeq 2.51$ is larger than $a_4^{1/4} \simeq 2.079$), the series (\ref{eq6.26}) does not exhibit a clear convergence. However, even if $a_n$ grows more like $(2.51)^n$, one still expects the $n=7$ contribution to be only $\pm \, (2.51/4)^7 = \pm \, 0.038$. In other words, one might hope that $a ( \mbox{$\frac{1}{4}$} ) = 0.144 \pm 0.040$, corresponding to $j_* = 4 \, [1-0.288(80) \, \nu + {\mathcal O} (\nu^2)]$. It will be interesting to see how this approximate prediction compares to GSF estimates. Note finally that the corresponding PN expansion for $a' ( \mbox{$\frac{1}{4}$} )$ has a worse convergence because (neglecting any logarithmic running)
\begin{equation}
\label{eq6.27}
a' ( \mbox{$\frac{1}{4}$} ) = \sum_{n \geq 3} \, n \, \frac{a_n}{4^{n-1}}
\end{equation}
includes a growing coefficient $n$. Because of the worsened convergence of (\ref{eq6.27}) it would be poorly justified to venture a numerical estimate of $a' ( \mbox{$\frac{1}{4}$} )$, beyond the fact that one expects it to be positive (on the basis of $a_3 + a_4$ giving the contribution $+ 1.54$ and $a_6 + a_7$ giving $+ 1.10$).

\section{About logarithmic terms in PN expansions}
\setcounter{equation}{0}

In the previous Sections we have considered, for simplicity, that the PN expansions of the various EOB functions (such as $A(u,\nu)$ or its $\nu$-linear contribution $a(u)$) contained only {\it powers} of the PN expansion parameter ($u=GM/c^2 \, r$ or $x = (GM \, \Omega / c^3)^{2/3}$ for functions of $x$). This simplifying assumption allowed us to connect our results to the recent results of the NR/EOB comparison which have all assumed that the function $A(u;\nu)$ admitted such power-law PN expansions, bacause they define $A(u)$ by Pad\'e-resumming the truncated Taylor expansion:
\begin{equation}
\label{eq7.1}
A^{\rm Taylor} (u) = 1-2u+\nu (a_3 \, u^3 + a_4 \, u^4 + a_5 \, u^5 + a_6 \, u^6 + \ldots) + {\mathcal O} (\nu^2) \, .
\end{equation}
We wish, however, to point out that the PN expansion of $A(u)$, contains logarithmic terms, starting at the 4PN level. This means that the coefficients $a_5 , a_6 , \ldots$ in Eq.~(\ref{eq7.1}) are not numerical constants but exhibit a ``logarithmic running'' with $u$, probably of the type
\begin{eqnarray}
\label{eq7.2}
a_5 (\ln \, u) &= &a_5^0 + a_5^1 \, \ln \, u \, , \nonumber \\
a_6 (\ln \, u) &= &a_6^0 + a_6^1 \, \ln \, u \, .
\end{eqnarray}
Indeed, we shall argue below that the logarithmic running of $a_5$ and $a_6$ is {\it linear} in $\ln \, u$. We, however, expect that some of the higher-order PN coefficients will exhibit some nonlinear runnings, say
\begin{equation}
\label{eq7.3}
a_n (\ln \, u) = a_n^0 + a_n^1 \, \ln \, u + \ldots + a_n^p (\ln \, u)^p \, ,
\end{equation}
where the maximum power $p$ increases (not faster than $n$) with $n$ (probably $p \leq n-4$).

\smallskip

Actually, as the $A(u)$ function is only related in a rather indirect (and ``non local'') way to the spacetime metric of a binary system, we shall discuss here the issue of logarithmic terms in the context of another function, defined along the sequence of {\it circular} orbits, which has been recently studied in detail within the GSF approach. We mean here Detweiler's gauge-invariant circular-orbit ``redshift'' function $u^t(y)$ \cite{Detweiler:2008ft}, where $y \equiv (Gm_2 \, \Omega / c^3)^{2/3}$ (which differs from the ``symmetric'' frequency parameter $x$ by the replacement $M \equiv m_1 + m_2 \to m_2$). Expanding $u^t(y)$ to first-order in the mass ratio $q \equiv m_1/m_2 \ll 1$ (so that $q=\nu + {\mathcal O} (\nu^2)$),
\begin{equation}
\label{eq7.4}
u^t(y) = (1-3y)^{-1/2} - q \, \bar u (y) + {\mathcal O} (q^2) \, ,
\end{equation}
defines a 1GSF-level function $\bar u (y)$ (here defined with a minus sign, so that $\bar u (y)$ be positive and involve mainly positive coefficients). Recently, Ref.~\cite{Blanchet:2009sd} has determined the 3PN-accurate post-Newtonian expansion of $\bar u (y)$ ($\equiv - u_{SF}^T (y)$ in their notation), namely
\begin{equation}
\label{eq7.5}
\bar u (y) = \bar u_1 \, y + \bar u_2 \, y^2 + \bar u_3 \, y^3 + \bar u_4 \, y^4 + {\mathcal O} (y^5) \, ,
\end{equation} 
where \cite{Detweiler:2008ft,Blanchet:2009sd}
\begin{equation}
\label{eq7.6}
\bar u_1 = 1 \, ; \quad \bar u_2 = 2 \, ; \quad \bar u_3 = 5 \, ; \quad \bar u_4 = \frac{121}{3} - \frac{41}{32} \, \pi^2 \, .
\end{equation}
Let us first point out that there seems to be a simple connection between the PN expansion coefficients $\bar u_n$ of $\bar u (y)$, and the PN expansion coefficients $a_n$ of the EOB function $a(u)$. Indeed, we can write
\begin{equation}
\label{eq7.7}
\bar u_n = a_n + k_n \, ,
\end{equation}
where $a_1 = 0  = a_2$ and the higher ones are given in Eq.~(\ref{eq4.31}), and where
\begin{equation}
\label{eq7.8}
k_1 = 1 \, ; \quad k_2 = 2 \, ; \quad k_3 = 3 \, ; \quad k_4 = 9 \, .
\end{equation}
It is remarkable that the difference $\bar u_n - a_n$ is an {\it integer}. [This might, however, be true only for the lowest PN orders, as $k_n$ might also depend on the lower-order $a_m$'s (with $m < n$).] Postponing to future work a detailed discussion of the link between $\bar u (y)$ and $a(u)$, we shall content ourselves here to note the simple connection (\ref{eq7.7}), (\ref{eq7.8}) and to heuristically argue that the presence of logarithmic terms in the PN expansion of $\bar u (y)$ will entail the presence of related logarithmic terms in the PN expansion of $a(u)$. 

\smallskip

We now focus on the presence of logarithmic terms in the PN expansion of $\bar u (y)$, which is more directly related to the PN expansion of the spacetime metric $g_{\mu\nu}$ of a binary system. The issue of logarithmic terms in PN expansions of $g_{\mu\nu}$ has been studied some time ago \cite{Anderson:1982fk,Blanchet:1985sp,Blanchet:1987wq}. Let us summarize here the main results. First, one should distinguish the issue of PN logarithms arising in the ``external metric'' (and also the ``wave zone'' metric), from that of PN logarithms arising in the ``inner metric'', or ``near zone'' metric. Logarithms enter the PN expansion of the external metric at the 3PN level \cite{Anderson:1982fk,Blanchet:1987wq}. They are linked to the cubic interaction (mass monopole) $\times$ (mass monopole) $\times$ (mass multipole) (see Eq.~(A6) of \cite{Blanchet:1987wq}). These terms enter the relation between the radiative multipole moments and the multipole moments of the source (and show up in the energy flux of binary systems at the 3PN level, see e.g. Eq.~(231) in the review \cite{Blanchet:2006zz}). Here, we are interested in the logarithms arising in the PN expansion of the inner (near-zone) metric. It was shown in \cite{Blanchet:1987wq} that the first logarithmic terms in the PN expansion of the inner metric arise at the 4PN level, and read (in a suitable gauge\footnote{Though some gauge transformations can be problematic (as discussed above for the Lorenz gauge), we have checked that the gauge transformations used in \cite{Blanchet:1987wq} do not affect the present computation.}), from Eq.~(6.39) there,
\begin{equation}
\label{eq7.9}
(\delta g_{00}^{\rm in})^{\ln c} = - \frac{8}{5} \, \frac{\ln \, c}{c^{10}} \, x^a \, x^b \, I(t) \, I_{ab}^{(6)} (t) \, ,
\end{equation}
where $I(t)$ is the mass monopole of the source (which can be approximated here as $I(t) \simeq M = m_1 + m_2$), and $I_{ab}^{(6)}$ the sixth time derivative of the quadrupole moment of the source (which can be approximated here by its Newtonian estimate $I_{ab} \simeq m_1 \, y_1^{\langle ab \rangle} + m_2 \, y_2^{\langle ab \rangle}$, where $y_1^a, y_2^a$ are the positions of the two masses, and where the angular brackets denote a symmetric trace-free (STF) projection). Remarkably, though the logarithmic term (\ref{eq7.9}) comes from a (hereditary) ``tail'' correction to the leading (Burke-Thorne) {\it radiation-reaction} contribution to the inner metric ($g_{00}^{{\rm rad}\mbox{-}{\rm reac}} = - \frac{2}{5} \, x^a \, x^b \, I_{ab}^{(5)} / c^7$, see \cite{Blanchet:1987wq}), the logarithmic term (\ref{eq7.9}) is symmetric under time reversal, and thereby survives as is in the conservative dynamics of the system. [This is due to the fact that the full {\it hereditary} radiation-reaction term that gives rise to (\ref{eq7.9}), is given by a ``logarithmic tail integral'' over the entire past (see Eq.~(6.38) of \cite{Blanchet:1987wq}) which is {\it time-asymmetric} without being time-odd as the leading radiation-reaction term.]

\smallskip

Let us now consider the contribution of the logarithmic term (\ref{eq7.9}) to Detweiler's redshift function (along the sequence of circular orbits of the ``small mass'' $m_1$)
\begin{equation}
\label{eq7.10}
u^t (y) = \left( \frac{dt}{ds} \right)_1 = \left( -g_{00}^1 - 2 \, g_{0i}^1 \, \frac{v_1^i}{c} - g_{ij}^1 \, \frac{v_1^i \, v_1^j}{c^2} \right)^{-1/2} \, ,
\end{equation}
where $g_{\mu\nu}^1 \equiv g_{\mu\nu}^{\rm reg} (y_1)$ denotes the (regularized) value of the spacetime metric at the location $y_1^i$ of the ``small mass'' $m_1$, and where $v_1^i$ denotes its coordinate velocity $dy_1^i / dt$.

\smallskip

We note first that the logarithmic term (\ref{eq7.9}) is easily seen to be proportional to the symmetric mass ratio $\nu$, and therefore to contribute directly to the $\bar u (y)$ piece within $u^t (y)$. There are, a priori, two sorts of contributions to $u^t(y)$ entailed by (\ref{eq7.9}), a direct one
\begin{equation}
\label{eq7.11}
\delta^{\rm direct} \, u^t (y) = \frac{1}{2} \, \delta \, g_{00}^{\rm in} \, (y_1)
\end{equation}
coming from the $-g_{00}^1$ contribution in (\ref{eq7.10}), and an indirect one coming from the fact that $u^t (y)$ should be computed for a fixed value of the orbital frequency $\Omega$, which implies that the coordinate radius $r_1$ of the orbit of $y_1^i$ must be perturbed by the additional term (\ref{eq7.9}) in the metric with respect to the value it had before taking this term into account. However, a simplification occurs in that the coefficient of the perturbation $\delta r_1$ of $r_1$ (approximately) vanishes. Indeed, the ``Newtonian'' contribution to the expansion of the R.H.S. of (\ref{eq7.10}) is $(r_{12} \equiv  \vert {\bm y}_{12} \vert , \ {\bm y}_{12} \equiv {\bm y}_1 - {\bm y}_2)$
\begin{equation}
\label{eq7.12}
\frac{Gm_2}{r_{12}} + \frac{1}{2} \, {\bm v}_1^2 = \frac{Gm_2}{r_{12}} + \frac{1}{2} \left( \frac{m_2}{M} \right)^2 \Omega^2 \, r_{12}^2 \, .
\end{equation}
In the limit $m_1 \ll m_2$, the $r_{12}$ (or $r_1 \simeq r_{12}$) derivative of (\ref{eq7.12}) vanishes by virtue of the Newtonian equations of motion, i.e. Kepler's law $\Omega^2 \, r_{12}^3 = GM$. [This also follows from the fact that we are expanding here the inverse of the Lagrangian of the particle ${\bm y}_1$.] This simplification means that Eq.~(\ref{eq7.11}) represents (in the limit $m_1 \ll m_2$) the full contribution to $-q \, \bar u (y)$ in Eq.~(\ref{eq7.4}). In other words, we have
\begin{equation}
\label{eq7.13}
q \, \delta^{\log} \, \bar u (y) = +\frac{4}{5} \, \frac{\ln \, c}{c^{10}} \, y_1^a \, y_1^b \, M \, \frac{d^6}{dt^6} \, (\mu \, y_{12}^{\langle ab \rangle}) \, .
\end{equation}

Evaluating the time derivative in (\ref{eq7.13}) for circular orbits (in the center of mass of the system, so that $I^{ab} = \mu \, y_{12}^{\langle ab \rangle}$ and ${\bm y}_1 = (m_2/M) \, {\bm y}_{12}$, as already used above), and noticing that $y = (Gm_2 \, \Omega / c^3)^{2/3}$ is proportional to $1/c^2$ so that $\ln \, y$ contains $-2 \, \ln \, c$, we find from (\ref{eq7.13}) that the leading logarithmic contribution to $\bar u (y)$ must be
\begin{equation}
\label{eq7.14}
\delta^{\log} \, \bar u (y) = + \frac{64}{5} \, y^5 \, \ln \, y \, .
\end{equation}

In other words, this is saying that the term formally denoted ${\mathcal O} (y^5)$ in Eq.~(\ref{eq7.5}) is of the form
\begin{equation}
\label{eq7.15}
(\bar u_5^0 + \bar u_5^1 \, \ln \, y) \, y^5
\end{equation}
with $\bar u_5^1 = 64/5 = 12.8$. Using the numerical data reported in Table~I of \cite{Blanchet:2009sd}, we have confirmed that this seems to be indeed the case with $\bar u_5^1 \simeq 12.8$ and  $\bar u_5^0 \simeq 114.4$.

\smallskip

By extending the arguments of \cite{Blanchet:1987wq} leading to (\ref{eq7.9}), it seems that the next logarithmic terms will be ${\mathcal O} (\ln \, c / c^{12})$, i.e. at the 5PN level, and involving only the first power of $\ln \, c$ instead of a possible $(\ln \, c)^2$ (which formally enters higher tail terms). This suggests that the next term is the PN expansion of $\bar u (y)$ will be of the form
\begin{equation}
\label{eq7.16}
(\bar u_6^0 + \bar u_6^1 \, \ln \, y) \, y^6 \, ,
\end{equation}
where $\bar u_6^1$ is an analytically calculable quantity. A fit of the numerical data of \cite{Blanchet:2009sd} suggests that $\bar u_6^1$ is comparable to $\bar u_5^1$ (i.e. $\sim 12$) and that $\bar u_6^0 \sim 360$. At higher PN levels, one expects higher powers of $\ln \, y$ to arise. The growth of the power of $\ln \, c$ with the iteration order was shown very generally in \cite{Blanchet:1985sp} and \cite{Blanchet:1987wq} (see Eq.~(5.2) there). See also Eq.~(5.3) in \cite{Blanchet:1987wq} which says that the power of $\ln \, c$ (say $p$) grows linearly with the PN order (say $N$): $p = N-c$. The constant $c$ was 3 in \cite{Blanchet:1987wq}, if we gauge the PN order by the power of $1/c$ in the spatial metric. The GSF data behind the figures given in \cite{Blanchet:2009sd} should be able to explore these phenomena. From the point of view of the present paper, which explores the contacts between the EOB formalism and GSF results, our main (tentative) conclusion (based part on the ``experimental link'' (\ref{eq7.7}) and part on the fact that the $A(u)$ function can be, in principle, computed from the spacetime metric) is the corresponding occurence of logarithmic terms in the EOB formalism, as sketched in Eqs.~(\ref{eq7.2}), (\ref{eq7.3}) above. We leave to future work a computation of the logarithmic coefficient $a_5^1$ in (\ref{eq7.2}).

\smallskip

As a final comment let us mention a possible subtlety. The logarithmic terms we have discussed above are {\it infrared} (IR) logs (linked to the matching between the near-zone and the wave-zone). There could also exist additional {\it ultraviolet} (UV) logs, linked to ``finite size effects'', i.e. to the matching between the ``body zones'' (of order the radii of the considered compact objetcs) and the near-zone. From an effective field theory point of view, logarithmic terms linked to finite-size effects would correspond to (logarithmic) {\it UV divergencies} in a pure point-mass description of a two-body system. Some time ago \cite{Damour:1982wm}, it was shown that the effective description of two gravitationaly compact bodies as two point masses (with action $S_{\rm point \, mass} = -\Sigma_A \int m_A \, ds_A$, together with analytic or dimensional regularization) is valid up to the {\it 5PN level}, where finite-size effects (linked to tidal interactions, and depending on the internal structure of the compact bodies via a certain ``relativistic Love number $k$''; see p.~83 of \cite{Damour:1982wm}) start appearing. The finite-size effects of body 1 correspond to the appearance of a quadrupole term in the metric, given (at the leading, Newtonian, approximation) by (see Eq.~(19) in Section~5 of \cite{Damour:1982wm})
\begin{equation}
\label{eq7.17}
\delta \, g_{00} \sim \frac{G^6}{c^{12}} \, k_1 \, m_1^5 \, m_2 \ \partial_{ij} \, \frac{1}{r_{12}} \ \partial_{ij} \, \frac{1}{\vert {\bm x} - {\bm y}_1 \vert} \, .
\end{equation}
[The label 1 on $k$ refers to body 1. We suppress here the unrelated index 2 on the Love number $k$  referring to its quadrupolar nature.] The additional finite-size effect (\ref{eq7.17}) can be described by augmenting the action of the two point masses by a non-minimal worldline coupling of the form $\Sigma_A \, \frac{1}{4} \, \mu_A \int ds_A \, {\mathcal E}_{\alpha\beta}^A \, {\mathcal E}^{A\alpha\beta}$, where ${\mathcal E}_{\alpha\beta}^A \equiv [u^{\mu} \, u^{\nu} \, C_{\mu\alpha\nu\beta}]^A$ is the ``electric'' worldline projection of the Weyl tensor \cite{Goldberger:2004jt,Damour:2009vw}. From an effective field theory point of view, one expects that the addition of such non-minimal couplings is needed not only to describe the finite part of extension effects, but also to ``renormalize'' the divergences ($1/\varepsilon$ poles in dimensional regularization) that arise when one tries to describe extended objects by a point-mass action. It is therefore a priori plausible that a dimensional-regularization (DR) calculation of the interaction of two point masses generate $1/\varepsilon$ poles at the 5PN level, linked to terms of the type (\ref{eq7.17}), but with a coefficient $\propto 1/\varepsilon$. As is well-known, $1/\varepsilon$ poles in DR correspond to logarithmic divergences, and always come accompanied by some logarithm of the ratio of the two relevant length scales: here that of the body zone, and the scale of variation of the metric near the considered body. This argument therefore suggests that the logarithm of (body size)/(scale of variation) $\sim GM/c^2 \, r_{12}$, i.e. $\ln \, u$ can arise at the 5PN level as an {\it UV log}. In other words, at the 5PN level (${\mathcal O} (u^6)$ in $g_{00}$) there might be two sources of $\ln \, u$: IR and UV. On the other hand, the recent work on tidal effects in neutron stars \cite{Damour:2009vw,Binnington:2009bb} has shown that the (quadrupolar) tidal coupling coefficient $\mu_A \propto k_A$ formally tended towards zero as the compactness of the neutron star tended towards that of a black hole. This may mean that the a priori expected 5PN finite-size divergencies of gravitationally interacting point masses cancell out, and do not give rise to $1/\varepsilon$ poles. In this case, there will be no UV source of $\ln \, u$ at the 5PN level, and the generalization of the 4PN IR argument above should give the entire 5PN logs.

\section{Conclusions}
\setcounter{equation}{0}

We have discussed various ways in which the computation of {\it conservative} Gravitational Self Force (GSF) effects on a point mass moving in a Schwarzschild background can inform us about some of the basic functions, $A(u;\nu)$, $\bar D (u;\nu)$, $Q(u,p_r)$, entering the Effective One Body (EOB) formalism. [Here $u \equiv G(m_1 + m_2) / r$ and $\nu \equiv m_1 m_2 / (m_1 + m_2)^2$.] Our main conclusions are the following:

\medskip

1. The recently published GSF calculation \cite{Barack:2009ey} of the ${\mathcal O} (\nu^1)$ shift of the orbital frequency of the Last Stable (circular) Orbit (LSO) gives us access to the combination (\ref{eq4.28}) of the values of $a(u) \equiv [\partial \, A (u;\nu) / \partial \, \nu]_{\nu = 0}$, Eq.~(\ref{eq3.16a}), and its first two derivatives, $a'(u)$ and $a'' (u)$, taken at the unperturbed LSO location $u_{\rm LSO} = \frac{1}{6} + {\mathcal O} (\nu)$.

\medskip

2. The current 3PN-based \cite{Damour:2000we,Damour:2001bu} knowledge of the beginning of the Taylor expansion of the function $a(u)$ ``explains'' $72.5\%$ of the GSF result. We expect that the missing $27.5\%$ will be contributed by the higher post-Newtonian (PN) contributions to the function $a(u)$, Eq.~(\ref{eq4.31}), and notably by $a_5$ (4PN contribution) and $a_6$ (5PN contribution). Combining the GSF result \cite{Barack:2009ey} and the 3PN-EOB results \cite{Damour:2000we,Damour:2001bu}, we determined in Eq~(\ref{eq4.40}) the value of a linear combination of $a_5 , a_6 , a_7$, etc$\ldots$ If the contributions of $a_7 , a_8$, etc$\ldots$ are relatively small (which can be argued for) this leads to determining a linear combination of $a_5$ and $a_6$, see Eqs.~(\ref{eq4.41}) or (\ref{eq4.42}).

\medskip

3. By combining the just mentionned approximate knowledge of a linear combination of $a_5$ and $a_6$ with the recent determination \cite{Damour:2009kr,Damour:2009ic} of the long and thin region of the $(a_5 , a_6)$ plane where the most accurate current EOB models \cite{Damour:2008gu,Damour:2009kr} exhibit an excellent agreement with NR data for comparable masses ($4\nu = {\mathcal O} (1)$), we found that they intersected near the point (\ref{eq4.45}) of the $(a_5 , a_6)$ plane. This suggests that the information coming from the $\nu \ll 1$ GSF study of the LSO is able to break the degeneracy among $a_5$ and $a_6$ left after tuning the two-parameter EOB $(a_5 , a_6)$ waveform to equal-mass ($\nu = \frac{1}{4}$) NR data \cite{Boyle:2007ft}. More work (both on the EOB side and on the GSF one) is however needed to confirm the tentative values (\ref{eq4.45}). [See the end of Section~4 for a more detailed discussion.]

\smallskip

Note that the values (\ref{eq4.45}), when inserted in the EOB formalism, predict the value of the LSO orbital frequency for all values of $\nu$, between the test-mass limit $\nu \ll 1$ and the equal-mass case $\nu = 0.25$. The exact EOB prediction is obtained by using the exact EOB results given in Section~4 (notably Eq.~(\ref{eq4.9})). Let us only indicate here that the final result for the $\nu$-dependence of the GSF $\cap$ NR-tuned $\hat\Omega_{\rm LSO} (\nu)$ can be approximately fitted by a quadratic polynomial in $\nu$ of the form
\begin{equation}
\label{eq8.1}
\hat\Omega_{\rm LSO} (\nu) \equiv G(m_1 + m_2) \, \Omega_{\rm LSO} (\nu) \simeq 6^{-3/2} [1+1.25 \, \nu + 1.87 \, \nu^2] \, .
\end{equation}

It should be noted on the result (\ref{eq8.1}) that the non-linear dependence on $\nu$ (here summarized by the term $+ \, 1.87 \, \nu^2$, but given in reality by a more complicated function $o_2 \, \nu^2 + o_3 \, \nu^3 + \ldots$) is numerically quite important for comparable-mass systems. E.g. in the equal-mass case ($\nu = \frac{1}{4}$) the 1GSF, $\nu$-linear result (\ref{eq2.17}), i.e. the contribution $+ \, 1.25 \, \nu$ in Eq.~(\ref{eq8.1}), predicts $\hat\Omega_{\rm LSO}^{\nu\mbox{-}{\rm lin}} = 6^{-3/2} [1.3125] \simeq 0.08930$, which is significantly (8\%) smaller than $\hat\Omega_{\rm LSO}^{\nu\mbox{-}{\rm quad}} = 6^{-3/2} [1.4294] \simeq 0.09726$, or the exact EOB value $\hat\Omega_{\rm LSO}^{\rm EOB} = 0.09670$. A good fit between EOB and equal-mass NR data requires that the equal-mass LSO frequency be roughly between 0.096 and 0.097. The approximate equation (\ref{eq8.1}) displays the needed {\it complementarity} between various approaches to the dynamics of binary systems: NR, GSF and EOB.

\medskip

4. We have also discussed (in Section~5) how the study of small-eccentricity orbits can allow one to confront the EOB formalism to GSF calculations. In particular, our Eq.~(\ref{eq5.21}) shows how the GSF computation, along the sequence of (quasi-)circular orbits, of the squared ratio between the radial (periastron to periastron) frequency and the azimuthal one gives us access to a function $\rho (x)$ of $x \equiv (G(m_1 + m_2) \, \Omega)^{2/3}$ which, when interpreted within the EOB formalism, is an $x$-dependent combination of $a(x)$, $a'(x)$, $a''(x)$ and $\bar d (x)$. Here, $a(u) \equiv [\partial \, A (u;\nu) / \partial \, \nu]_{\nu = 0}$ as above, and $\bar d (u) \equiv [\partial \, \bar D (u;\nu) / \partial \, \nu]_{\nu = 0}$, where the EOB metric function $\bar D (r;\nu)$ is defined in Eq.~(\ref{eq3.13}). More precisely, we recommend to study not only (as in \cite{Barack:2009ey}) the behaviour of $\rho (x)$ near the LSO (where $(\omega_r / \Omega)^2$ vanishes), but also below the LSO (for $\frac{1}{6} < x < \frac{1}{3}$), and around $x=0$. We have explicitly given the first two terms (2PN and 3PN) in the Taylor expansion (\ref{eq5.32}) of $\rho (x)$ near $x=0$. See Eq.~(\ref{eq5.35}) (which also follows from results given in \cite{Damour:1999cr}). We emphasize the need of this comparison between 3PN results and GSF ones for confirming the validity of the results of \cite{Barack:2009ey} (similarly to the way 3PN results were recently used \cite{Blanchet:2009sd} to confirm the validity of the gauge-invariant GSF result of \cite{Detweiler:2008ft}). We have also explicitly shown how the determination of the higher-order terms ($\rho_4 \, x^4 + \rho_5 \, x^5 + \ldots$, corresponding to 4PN $+$ 5PN $+ \, \ldots$) in the Taylor expansion of $\rho (x)$ can give us access to specific combinations of the higher PN coefficients $a_5 , a_6 , \ldots ; \bar d_4 , \bar d_5 , \ldots$ entering the PN expansions of the two functions $a(u)$ and $\bar d (u)$, see Eq.~(\ref{eq5.33a}--\ref{eq5.33e}) and (\ref{eq5.41}). In particular if $10 \, a_5$ {\it dominates} over $\bar d_4$, this could give us some interesting confirmation of the tentative determination (\ref{eq4.45}) of $a_5$.

\medskip

5. We have also discussed several other ways to confront (conservative) GSF calculations to the EOB formalism. Some of them present challenges to the GSF line of work: such as the GSF determination of the total conserved energy and angular momentum (using either the {\it second} GSF approximation, or a detailed study of unbound orbits; see beginning of Section~6). An easier way of getting new, quantitative information from GSF studies is to study the special {\it zero-binding zoom-whirl} motion which starts, in the infinite past, with zero kinetic energy at infinite separation, and ends up, in the infinite future, whirling indefinitely around some limiting finite separation. We showed how the GSF study of this special motion could give us access to (at least) {\it three} dimensionless observables: $j_*$, $\hat\Omega_{\rm whirl}$ and $(\omega_r^2 / \Omega^2)_{\rm whirl}$. We related these observables to the values of various ($\nu$-linearized) EOB functions at $u = 1/4$. We also ventured an approximate prediction for the value of $a ( \mbox{$\frac{1}{4}$} )$ entering the EOB prediction for $j_*$, see Eqs.~(\ref{eq6.22}) and (\ref{eq6.26}).

\medskip

6. In addition, we have shown that the logarithmic terms $\ln \, c$ that appear, at the 4PN level, in the post-Newtonian expansion of the near-zone metric of a gravitating system \cite{Blanchet:1987wq} give rise to a (Renormalization-Group-type) logarithmic running of the expansion coefficients of various functions: the ``redshift'' function $u^t (y)$ of Ref.~\cite{Detweiler:2008ft} (for which we computed the leading logarithmic term, Eq.~(\ref{eq7.14})), and, arguably, the EOB function $a(u)$ (and the other EOB functions, such as $\bar d (u)$). We leave to future work an exploration of the effects of including such logarithmic terms, as in Eq.~(\ref{eq7.2}), in the EOB formalism. Though this will affect the details of our GSF/EOB comparison, we do not expect that this will introduce drastic changes because the EOB/NR comparison (and in paricular the determination of the ``good fit'' region in the $(a_5 , a_6)$ plane) mainly depends on $a_5$ and $a_6$ as {\it effective parameters}, describing the shape of the $A(u)$ function in an interval between the LSO and the EOB light-ring (i.e. roughly between $u \sim 1/6$ and $u \sim 1/3$). Therefore, if we consider ``running parameters'', the EOB/NR comparison will be mainly sensitive to $a_5^{\rm effective} = \langle a_5 (\ln \, u) \rangle$ and $a_6^{\rm effective} = \langle a_6 (\ln \, u)\rangle$, where the brackets denote an operation of averaging over an interval of $u$ close to the LSO. As a rough approximation we expect that $a_5^{\rm effective} \simeq a_5 (\ln \, u_{\rm LSO})$ and $a_6^{\rm effective} \simeq a_6 (\ln \, u_{\rm LSO})$. Therefore, as the GSF results depended on the behaviour of the $a(u)$ function near the (unperturbed) LSO, the EOB/NR $a_5^{\rm effective}$, $a_6^{\rm effective}$ should be also approximately relevant in the EOB/GSF comparison.

\bigskip

Finally, let us remark that:

\smallskip

a. We recommend that the GSF studies based on the use of the Lorenz gauge be systematically reformulated (or at least re-expressed) in terms of an ``asymptotically flat'' coordinate system. Indeed, the present use of a ``non-asymptotically flat'' coordinate system is not only confusing for general physicists, but can cause real errors (e.g. when considering inspiralling motions where the ``renormalization'' factor $1+\alpha$, Eq.~(\ref{eq2.8}), connecting $t_{\rm Lorenz}$ to $t_{\rm flat}$, would become adiabatically time-dependent).

\smallskip

b. Our result (\ref{eq4.45}) on the way the GSF result \cite{Barack:2009ey} breaks the degeneracy of the EOB-NR constraint on $(a_5 , a_6)$ implies, in particular, that the {\it nonlinear dependence on} $\nu$ entailed by the Pad\'e-resummed definition of the $A(u;\nu)$ function within the EOB formalism \cite{Damour:2000we} plays an important r\^ole. This can be seen, for instance, in considering an $A$ function of the type
\begin{equation}
\label{eq8.2}
A_P^{\nu\mbox{-}{\rm linear}} (u;\nu) = 1-2u + \nu \, a_P (u; a_5 , a_6) \, ,
\end{equation}
where $a_P (u,a_5 , a_6)$ is (uniquely) {\it defined} by requiring that the equal-mass value $A_P^{\nu\mbox{-}{\rm linear}} (u;1/4)$ be equal to the normal Pad\'e-resumed $A_{\mbox{\footnotesize Pad\'e}}^{\rm EOB} (u;\nu;a_5,a_6)$ (whose structure was recalled in Section~4, before Eq.~(\ref{eq4.43})). Then, one finds that, all along the banana-like ``good fit'' region in the $(a_5 , a_6)$ plane \cite{Damour:2009kr}, the fractional $\nu$-derivative (at $\nu = 0$) $c_{\Omega}$, Eq.~(\ref{eq2.16}), of $GM \, \Omega_{\rm LSO}$ predicted by the $\nu$-linear $A$ function (\ref{eq8.2}) stays in the vicinity of $c_{\Omega} [A^{\nu\mbox{-}{\rm linear}} ] \simeq 0.82$, practically independently of $(a_5 , a_6)$, when varying them along the center of the banana-like ``good EOB/NR fit'' region in the $(a_5 , a_6)$ plane. Note that the value 0.82 is only $65.6\%$ of the GSF value (\ref{eq2.17}). By contrast, the value of $c_{\Omega}$ corresponding to the exact (Pad\'e) EOB function $A_{\mbox{\footnotesize Pad\'e}}^{\rm EOB} (u;\nu;a_5,a_6)$ does vary along the central line of the ``good fit'' $(a_5 , a_6)$ region, and does reach the GSF value (\ref{eq2.17}) at (and only at) the particular values (\ref{eq4.45}). It would be interesting to confirm the need of such nonlinear behaviour in $A(u;\nu)$ by exploring in more detail than has been done so far the EOB/NR comparison for several mass ratios.

\end{document}